\documentclass[fleqn,usenatbib]{mnras}

\usepackage{newtxtext,newtxmath}

\usepackage[T1]{fontenc}
\usepackage{ae,aecompl}
\usepackage{rotating}
\usepackage{color}

\usepackage{graphicx}	
\usepackage[utf8]{inputenc}
\usepackage[dvipsnames]{xcolor}
\definecolor{royalfuchsia}{rgb}{0.79, 0.17, 0.57}

\title{Optical Counterparts of ULXs in NGC 1672}

\author[S. Allak et al.]{S. Allak,$^{1,3}$\thanks{E-mail:0417allaksinan@gmail.com}
A. Akyuz,$^{2,3}$ 
E. Sonbas,$^{4,5}$
K. S. Dhuga$^{5}$
\\
$^1$Department of Physics, University of Çanakkale Onsekiz Mart, 17100, Çanakkale, Turkey \\
$^2$Department of Physics, University of Çukurova, 01330, Adana, Turkey\\
$^3$Space Science and Solar Energy Research and Application Center (UZAYMER), University of Çukurova, 01330, Adana, Turkey\\
$^4$Adiyaman University, Department of Physics, 02040 Adiyaman, Turkey\\
$^5$Department of Physics, The George Washington University, Washington, DC 20052, USA\\
}
\date{Accepted XXX. Received YYY; in original form ZZZ}

\pubyear{2020}

\begin{document}
\label{firstpage}
\pagerange{\pageref{firstpage}--\pageref{lastpage}}
\maketitle

\begin{abstract}
In this work, we deploy archival data from {\it HST}, {\it Chandra}, {\it XMM-Newton}, and {\it Swift-XRT}, to probe the nature of 9 candidate ULXs in NGC 1672. Specifically, our study focuses on using the precise source positions obtained via improved astrometry based on {\it Chandra} and {\it HST} observations to search for and identify potential optical counterparts for these ULXs. Unique optical counterparts are identified for two of the ULX candidates i.e., X2 and X6; for three of the candidates i.e., X1, X5 and X7, we found two potential counterparts for each source within the respective error radii. No optical counterparts were found for the remaining four sources. The spectral energy distribution of X2 is fitted to a blackbody spectrum with a temperature of $\sim$ 10$^{4}$K and the spectral class of the source is determined to be B7$-$A3, a supergiant donor star. We used colour magnitude diagrams (CMDs) to investigate ages of the counterparts. Of all the sources studied, X9 exhibits the most variability whereby the X-ray flux varies by a factor of $\sim$ 50 over a time period spanning 2004 to 2019, and also traces a partial q-curve-like feature in the hardness-intensity diagram, hinting at possible spectral transitions.
\end{abstract}


\section{Introduction}

Ultraluminous X-ray sources (ULXs) are non-nuclear point-like sources in a number of external galaxies. Their X-ray luminosities are above a threshold luminosity of L$_{X}$ > 10$^{39}$ erg s$^{-1}$, exceeding the Eddington limit for a typical 10 M$_{\odot}$ stellar-remnant black hole (see review by \citealp{2017ARA&A..55..303K, 2021AstBu..76....6F}).

Several possibilities as to the nature of ULXs continue to be discussed in the literature: (a) an early model, though now seemingly less likely, poses the existence of BHs in the intermediate mass range of M $\sim$ 10$^{2}$- 10$^{5}$ M$_{\odot}$, accreting at sub-Eddington rates \citep{1999ApJ...519...89C,2004ApJ...614L.117M}, (b) current models, based on recent data, tend to lean toward stellar-mass compact objects with a possible
combination of effects such as geometric beaming, and/or accretion at super-Eddington limits \citep{2002ApJ...568L..97B,2007Ap&SS.311..203R, 2007MNRAS.377.1187P,2009MNRAS.393L..41K,2018ApJ...857L...3W}. Indeed, the recent detection of pulsations in a handful of ULX sources strongly argues in favor of at least a fraction of these sources hosting neutron stars (NSs) \citep{2014Natur.514..202B,2016ApJ...831L..14F,2017Sci...355..817I,2017MNRAS.466L..48I,2018MNRAS.476L..45C,2019MNRAS.488L..35S,2020ApJ...895...60R}.

With the current total of NS ULXs (with observed pulsations) standing at about a half-dozen, the question of the relative proportion of ULXs hosting a NS as opposed to a stellar-mass BH, has become a hotly debated topic \citep{2016MNRAS.458L..10K,2017MNRAS.470L..69M, 2019ApJ...875...53W}. In addition to the signature of coherent pulsations, NS ULXs also exhibit a large dynamic range in variability in their observed light curves. This feature has been suggested as a possible method of identifying NS ULXs in the absence of pulsations \citep{2018MNRAS.476.4272E}. Indeed, a recent {\it Swift-XRT} monitoring campaign of M51 \citep{2020ApJ...895..127B} has yielded evidence for counterpart transient ULXs through the observation of long-term variability in their light curves. Likewise, the presence of a cyclotron resonance scattering feature in the X-ray spectrum of ULX-8 in M51, discovered by \cite{2019MNRAS.486....2M}, provides strong evidence for the nature of the compact object i.e., a NS, and a measure of the associated magnetic field.

In the present work, we focus on NGC 1672, a late-type barred spiral galaxy at a distance of 16.3 Mpc \citep{2000AJ....119..612D}. This galaxy has been studied by \cite{2011ApJ...734...33J}, who identified 28 X-ray sources within the D$_{25}$ region of the galaxy. Among these sources, 9 were observed to have X-ray luminosities consistent with those associated with ULXs. Our primary goal is to search for and identify potential optical counterparts for these 9 candidate ULXs by deploying all available optical data. In addition, a secondary goal is to use the available {\it Chandra}, {\it XMM-Newton} and {\it Swift} archival data, to study the X-ray spectral and temporal variations of the candidate ULXs.

Optical studies provide valuable information regarding the nature of the donor star, disk geometry, and can place constraints on the mass of the accretor. Technically, the optical emission observed in ULX binaries can be due to the accretion disk, the donor star, and/or some combination of both. Many recent studies \citep{2007MNRAS.376.1407C,2008MNRAS.386..543P,2011ApJ...737...81T,2012ApJ...745..123G,2012MNRAS.420.3599S,2014MNRAS.444.2415S,2018MNRAS.480.4918A,2019ApJ...884L...3Y}, focusing on optical variability, multi-band colors, and SED modeling, strongly suggest that the optical emission is likely contaminated or even dominated by reprocessed radiation from an irradiated accretion disk.

Unlike X-ray studies, the number of ULXs defined by optical observations is still limited since they are quite faint in the optical band (m$_{V}$ > 21 mag.) and tend to be located in crowded star-forming regions. Therefore, identification of counterparts often requires the use of the Hubble Space Telescope ({\it HST}\footnote{https://archive.stsci.edu/hst/search.php}) for optical imaging. In addition, precise astrometry between high resolution {\it Chandra} X-ray and {\it HST} images is also essential \citep{2010MNRAS.403L..69P,2011ApJ...734...23G, 2013MNRAS.432..506P, 2017MNRAS.467L...1E, 2018MNRAS.475.3561U,2022MNRAS.510.4355A}. Identification of point-like (potential) counterparts of ULXs with blue colors are indicative of early-type, OB stars. This of course goes with the caveat that the observed blue color may result from dominant contributions from the X-ray irradiation of the accretion disk and/or the companion star facing the X-ray source \citep{2010MNRAS.403L..69P,2012ApJ...758...28J,2018ApJ...854..176V}. As a rare example, \cite{2014Natur.514..198M}
reported that photospheric absorption lines have been detected from the donor star in the blue part of the spectrum of P13 in NGC 7793. On the other hand, the systematic search for nearby ULX counterparts in the near-infrared (H band) revealed that they might be red supergiants \citep{2014MNRAS.442.1054H,2016MNRAS.459..771H,2017MNRAS.469..671L,2020MNRAS.497..917L}. In these systems the contribution of the accretion disk is expected to be lower in the near-infrared than in the optical band and additionally, the irradiation of the donors is unlikely to be a significant effect on the observed emission because of the large separations of the companions \citep{2007MNRAS.376.1407C,2015MNRAS.453.3510H}.\\

The layout of the paper is as follows: in section \ref{sec:2}, we describe the data selection and the reduction procedures for both the optical and X-ray analysis, including the steps needed to perform astrometry, construct spectral energy distributions, hardness-intensity diagrams, color magnitude diagrams, and determine the spectral optical index. In section \ref{sec:3}, we present our results of the optical analysis, astrometry, and X-ray spectral fits and temporal variability studies, as well as, the outcomes for particular sources. Finally, we summarize and conclude our main findings in section \ref{sec:4}.

\section{Data Reduction and Methodology} \label{sec:2}
\subsection{Optical}
\subsubsection{Astrometry}

NGC 1672 was observed by {\it HST} in 2005 and 2019 using the ACS/WFC (Advanced Camera for Surveys/Wide Field Channel) and WFC3/UVIS (The Wide Field Camera 3), respectively. Details of the observations are given in Table \ref{T:observations}.

\begin{figure*}
\begin{center}
\includegraphics[angle=0,scale=0.27]{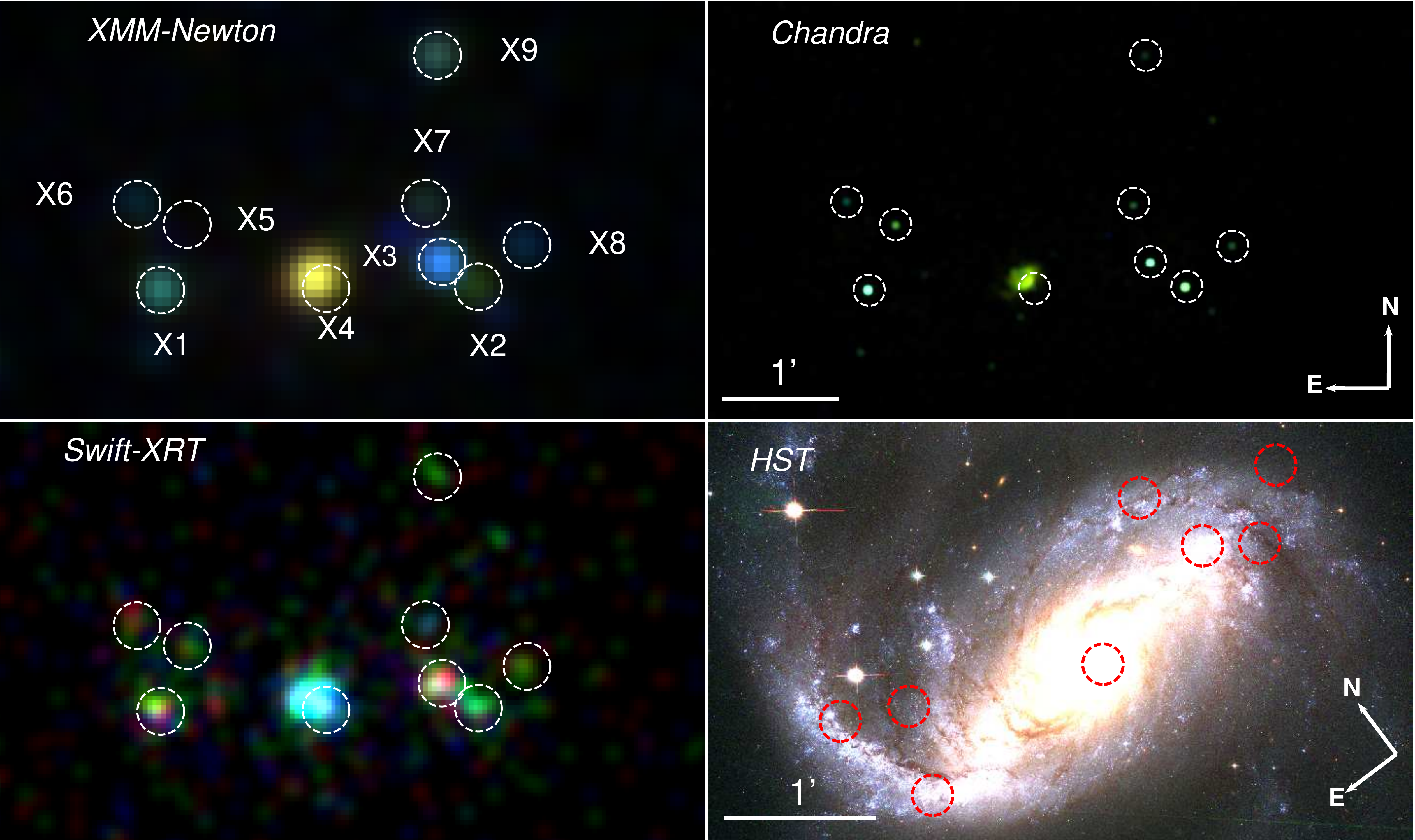}
\caption{Three-color X-ray (upper left {\it XMM-Newton}, upper right {\it Chandra} and lower left {\it Swift-XRT}) and {\it HST} (lower right) images of the NGC 1672 galaxy. For X-ray images, red, green and blue (RGB) represent 0.3–1 keV, 1–2.5 keV and 2.5–8 keV emissions, respectively and images smoothed with a 5$\arcsec$ Gaussian. The ULX candidates are indicated with dashed white circles. Red circles represent the positions of 8 ULXs on the {\it HST} RGB image (R:F814W, G:F550M and B:F435W). The X9
is not in the field of view of the {\it HST} images. All X-ray images are the same scale.}
\label{F:1}
\end{center}
\end{figure*}

We note at the outset that our numbering scheme for the sources differs from that used by \cite{2011ApJ...734...33J}; we label the candidate ULXs according to their increasing {\it Chandra} counts i.e., X1 (\#25), X2 (\#4), X3 (\#5), X4 (\#13), X5 (\#24), X6 (\#28), X7 (\#7), X8 (\#1) and X9 (\#6). The numbers in parentheses correspond to the source numbers in \cite{2011ApJ...734...33J}. We also mention that \cite{2011ApJS..192...10L} identified four ULXs corresponding to our X1, X2, X3 and X4. Figure \ref{F:1} shows the location of the 9 ULX candidates on three-color (red, green and blue; RGB) images comprising the data from {\it XMM-Newton}, {\it Chandra}, {\it Swift-XRT} and {\it HST}.

Determining the optical counterparts of the ULXs requires accurate source positions provided by astrometric corrections. {\it Chandra} and {\it HST} images were used to obtain astrometric precision by following the method we used in our previous studies \cite{2022MNRAS.510.4355A} (and references therein). We chose deep {\it Chandra} ACIS-S (ObsID 5932) and {\it HST} ACS (ObsID j6n202010) images for astrometric corrections of the sources. We used {\it wavedetect} tool in {\scshape ciao} and {\it daofind} tool in {\scshape iraf} for source detection in both images, respectively. We found four reference point sources between {\it Chandra} and {\it HST} images. All of the matched sources are located on ACIS-S with a moderate offset from the optical axis in {\it Chandra} data. Properties of the reference sources are given in Table \ref{T:astrometry}. The offsets between the reference sources were calculated with 95\% confidence level. The astrometric errors between the {\it Chandra} and {\it HST} images were determined as 0\farcs08 $\pm$ 0.03 in R.A. and 0\farcs07 $\pm$ 0.03 in Dec. As a result, we found the positions of the optical counterparts of ULXs on the {\it HST} image within an error radius of 0\farcs21 at 95\% confidence level. The corrected {\it HST} coordinates of the ULXs are also given in Table \ref{T:astrometry}. We searched for optical counterparts of all ULXs within their respective error radii.

We found unique optical counterparts for X2 ({\it X2$\_1$}) and X6 ({\it X6$\_1$}). Two potential counterparts each were found for X1 ({\it {\it X1$\_1$}} and {\it X1$\_2$}), X5 ({\it X5$\_1$} and {\it X5$\_2$} ) and X7 ({\it X7$\_1$} and {\it X7$\_2$}) respectively. The labels in parentheses indicate the corresponding optical counterpart. However, we could not identify any optical counterpart(s) for X3, X4 and X8 within their respective error radii: X9, unfortunately, was not observed by {\it HST}. The candidate optical counterparts are marked on the ACS/WFC/F814W image displayed in Figure \ref{F:can}.

\subsubsection{Photometry}

Point-Spread Function (PSF) photometry was performed with {\scshape dolphot} v2.0 \citep{2000PASP..112.1383D} to determine
the magnitudes of optical sources using the {\it HST} data. The {\scshape acsmask} task was used to remove pixels flagged as bad in images. For the next step, the tasks {\scshape splitgroups} and {\scshape calcsky} were run to split the image files into each chip and to create the sky background for each image, respectively. We then derived the magnitudes of sources using the set of parameters for the ACS/WFC and WFC3/UVIS recommended by {\scshape dolphot} user’s guide. Obtained magnitude values were corrected with Galactic extinction A$_{V}$ = 0.065 mag.\citep{2011ApJ...737..103S}.

The dereddened Vega magnitudes of optical counterparts are given in Table \ref{T:fotometri}. We calculated the absolute magnitudes, M$_{V}$, for both UVIS/F555W (V) and WFC/F550M ($\simeq$V) with adopted distance of 16.3 Mpc. These values are given in Table \ref{T:tab6}. In addition, the calculated color indices to determine the spectral type of optical counterparts of ULXs are also given in Table \ref{T:tab6}.

\subsubsection{Spectral energy distribution} \label{sec:sed}

Spectral energy distribution (SEDs) of the optical counterparts have been constructed to obtain the spectral characteristics of ULXs using the derived flux values given in Table \ref{T:fotometri}. The wavelengths of the filters are selected as the pivot wavelength, obtained from {\scshape pysynphot}\footnote{https://pysynphot.readthedocs.io/en/latest/}, in SED plots. The excess of the flux contaminated
by continuum emission within the F658N filter is clearly visible compared to other filters. We applied the procedure as given by \cite{2011ApJ...734...33J} to extract the nebular emission. Narrow band F658N image covers two adjacent lines such as [NII] ($\lambda$6548 and [NII] ($\lambda$6583). For this, calibration was performed between the F550M and F814W images using the flux values of ten stars which are in a radius of 8 pixels. Then, we built a normalized image from the weighted average of F814W and F550M. The resulting image was subtracted from the F658N image. As a result, we derived the continuum subtracted flux as 1.01$\times$10$^{-18}$ erg cm$^{-2}$ s$^{-1}$ \AA$^{-1}$ from the circular aperture of radius 0.$\arcsec$21.\\

\begin{figure}
\begin{center}
\includegraphics[width=\columnwidth]{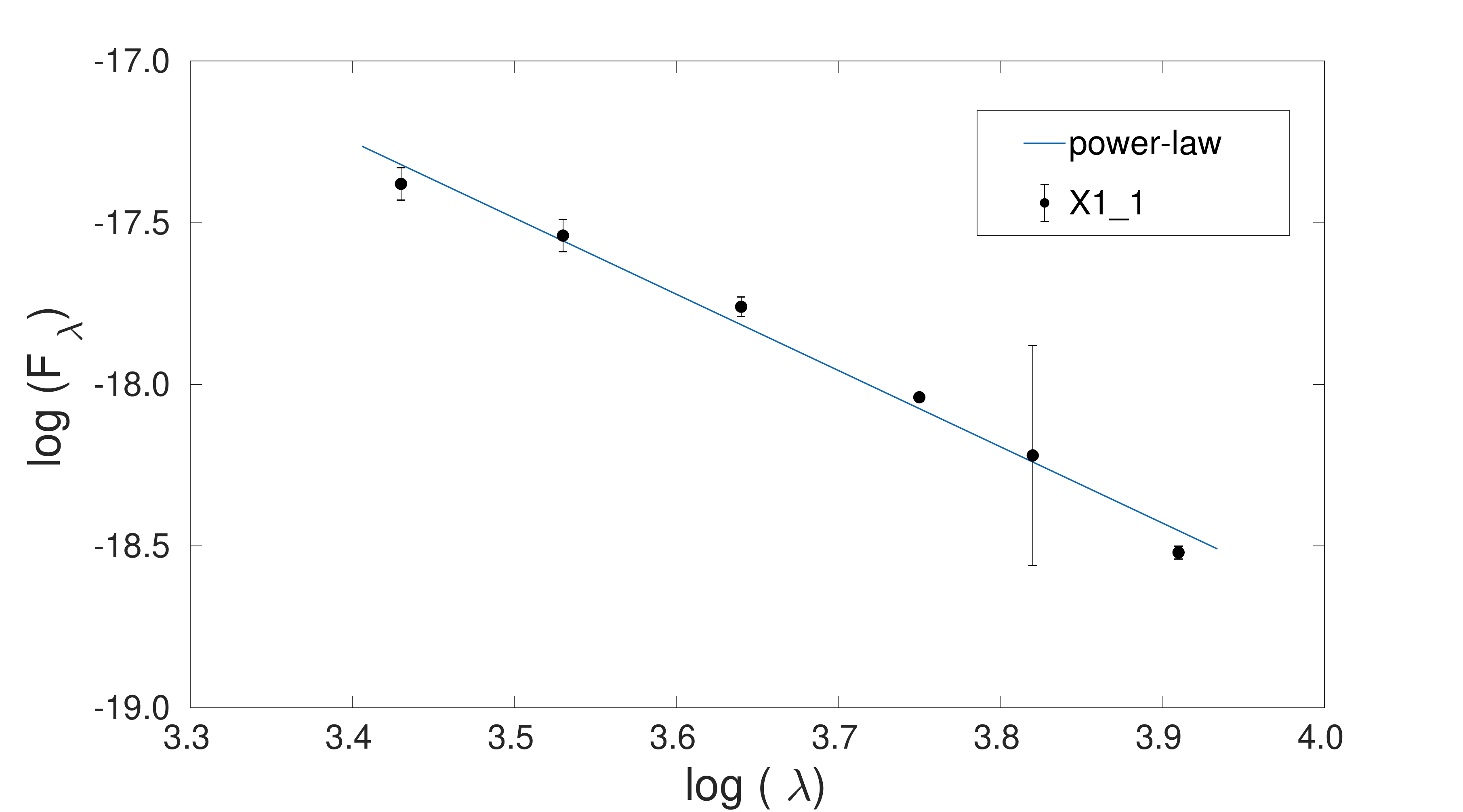}
\caption{The reddening corrected SED of {\it X1$\_1$}. The {\it power-law} model is shown by blue solid line. All data are shown with dark circles and their
respective errors with bars. The {\it power-law} model has $\alpha$ $=$ -2.13 $\pm$ 0.16. The units of y and x axes are erg s$^{-1}$ cm$^{-2}$ \AA$^{-1}$ and \AA, respectively.}
\label{F:SEDsX1}
\end{center}
\end{figure}

\begin{figure}
\begin{center}
\includegraphics[width=\columnwidth]{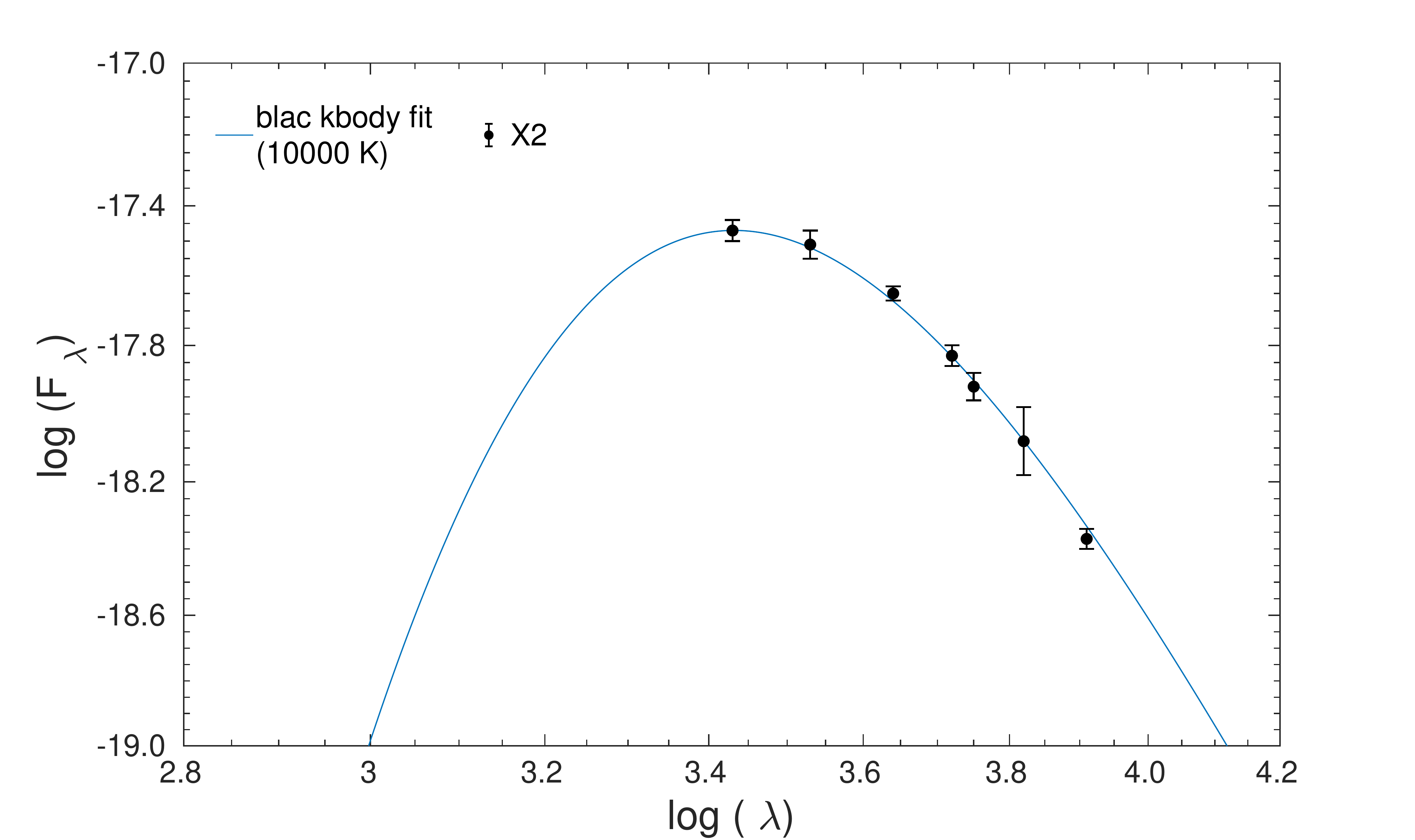}
\caption{The reddening corrected SED of {\it X2$\_1$}. The {\it blackbody} model is shown by blue solid line. All data are shown with dark circles and their respective errors with bars. The {\it blackbody} has a temperature of $\sim$ 10$^{4}$ K. The units of y and x axes are erg s$^{-1}$ cm$^{-2}$ \AA$^{-1}$ and \AA, respectively.}
\label{F:SEDsX2}
\end{center}
\end{figure}

We attempt to constrain the nature of companions by fitting SEDs with a {\it blackbody} or a {\it power-law} (F $\propto$ $\lambda^{\alpha}$) spectrum. To obtain a {\it blackbody} spectrum, a publicly available code was used with {\scshape optimset} and {\scshape fminsearch} functions in {\scshape matlab}\footnote{https://www.mathworks.com/matlabcentral/fileexchange/20129-fit-blackbody-equation-to-spectrum}. We found that SED fitting was only possible when data from more than three filter bands were available. On the other hand, of the 8 counterparts identified, 4 are observed in only more than three filter band (see Table \ref{T:fotometri}). The SED of {\it X1$\_1$} is adequately fitted by a {\it power-law} spectrum with $\alpha$ $=$ -2.13 $\pm$ 0.16. The {\it X2$\_1$} SED is well fitted by a {\it blackbody} spectrum with a temperature of 10$^{4}$ $\pm$ 113 Kelvin (K) at 95\% confidence level. This temperature corresponds to a late-type B supergiant donor \citep{1981Ap&SS..80..353S}. Additional extinction values were used to check for the possibility that SED would change to a {\it power-law} shape. For this, we increased the Galactic extinction value by two to ten times, however, an acceptable fit for the {\it power-law} shape could not be obtained. Moreover, we also used CK04 standard stellar spectra templates \citep{2004A&A...419..725C} to confirm spectral type of the {\it X2$\_1$}. The CK04 templates were derived with metallicity of [Fe/H]$=$-1.74 \citep{2020AstBu..75..384T} and extinction of A$_{V}$ $=$ 0.065 mag. Also, synthetic spectra were normalised with Vega magnitude (m$_{v}$ = 0). It was found that {\it X2$\_1$} is compatible with the SED of a late-type B star. The reddening corrected SEDs of {\it X1$\_1$} and {\it X2$\_1$} are shown in Figure \ref{F:SEDsX1} and \ref{F:SEDsX2}, respectively. No acceptable model-fits were obtained for the remaining optical counterparts.

\subsubsection{Color-magnitude diagrams}

\begin{figure}
\begin{center}
\includegraphics[width=\columnwidth]{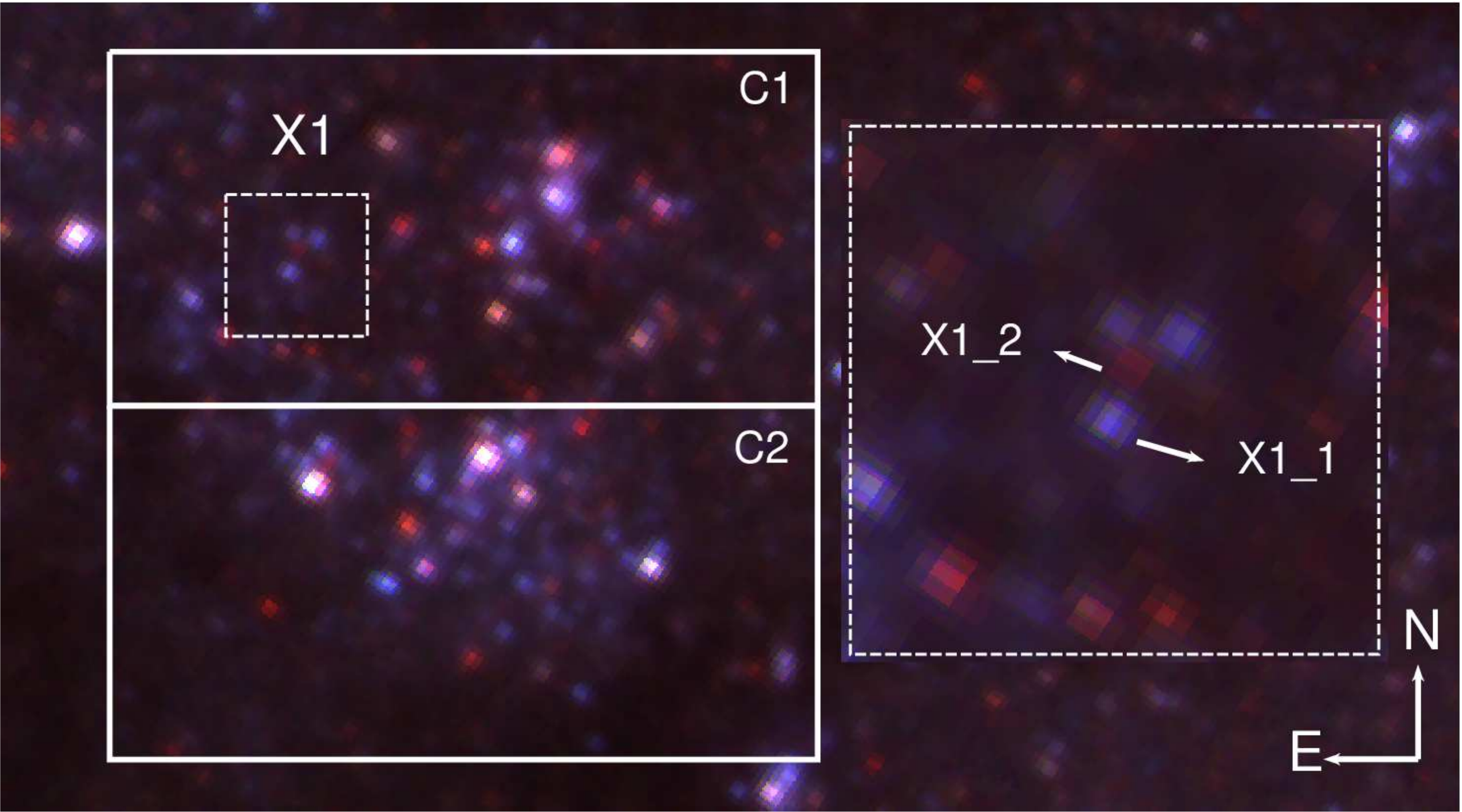}
\includegraphics[width=\columnwidth]{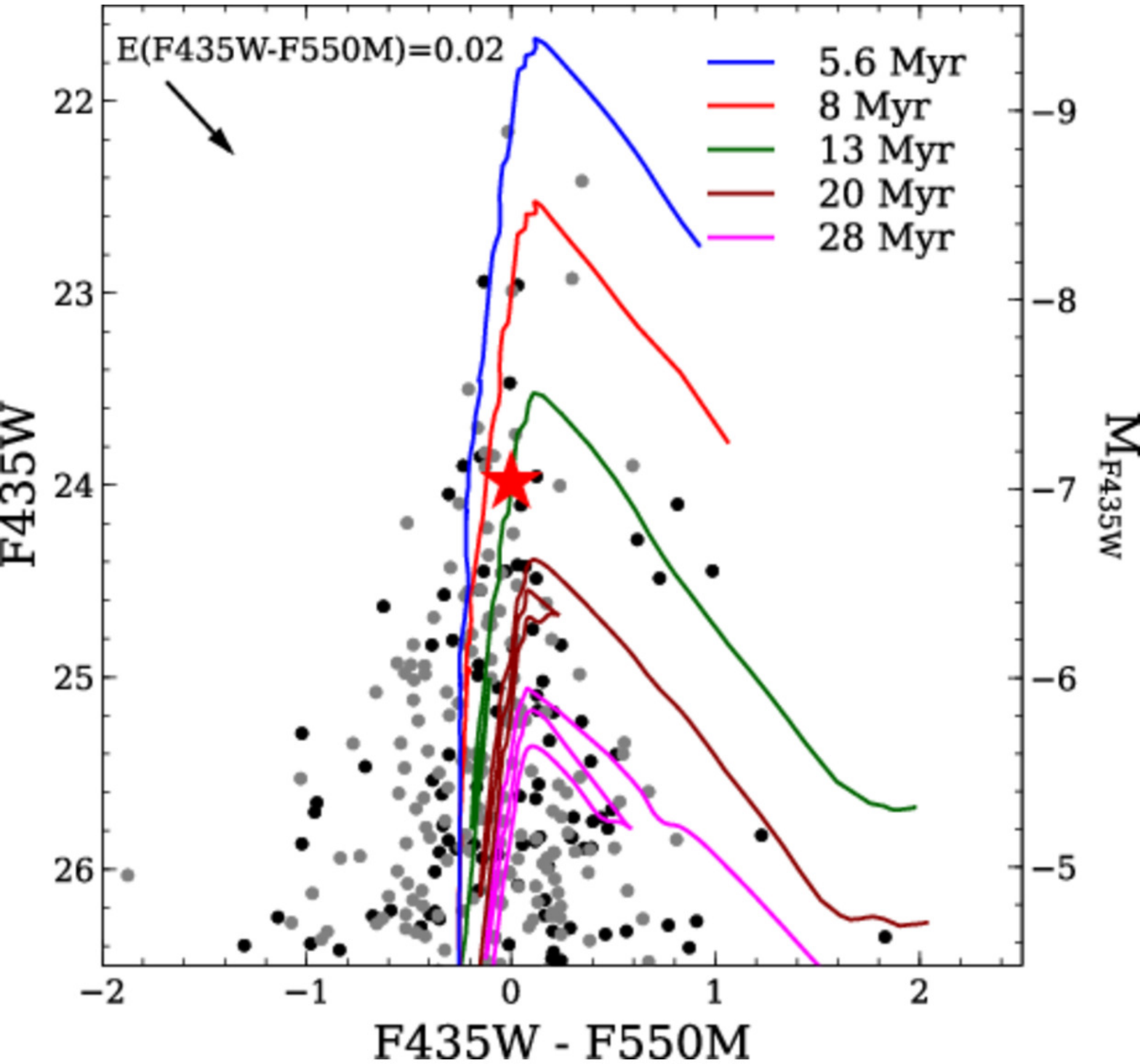}
\caption{Upper: Optical counterparts of X1 and their environments are shown on the {\it HST} RGB image. Two large groups of stars are indicated by two white rectangles (C1 and C2) with the size of (2.$\arcsec$5$\times$5$\arcsec$) on the image. Zoomed view of the optical counterparts for clarity is given upper right. Lower: CMD for the {\it X1$\_1$}, two group of stars and field stars around {\it X1$\_1$}. The red star, black and gray dots represent the {\it X1$\_1$}, field stars within C1 and C2, respectively. The isochrones have been corrected for extinction of A$_{V}$ $=$ 0.065 mag and the black arrow shows the reddening line.}
\label{F:CMD1}
\end{center}
\end{figure}

In order to investigate the relation of the optical counterparts with their environments and estimate their ages, the color-magnitude diagrams (CMDs) were obtained assuming the optical emission of the donor stars dominate.
Two CMDs, F435W (B) versus F435W$-$F550M (B-V) and F550M (V) versus F550M$-$F814W (V-I), were derived for optical counterparts and the field stars. The metallicity of [Fe/H]$=$-1.74 was used to obtain the corresponding PARSEC \citep{2012MNRAS.427..127B} isochrones. The distance modulus was calculated as 31.06 magnitude using the adopted distance of 16.3 Mpc. The reddening in the direction of the galaxy was calculated as E(F435W$-$F550M)$=$0.02. CMDs were obtained for {\it X1$\_1$}, {\it X2$\_1$}, {\it X5$\_1$} and {\it X7$\_1$}; the remaining sources were not detected in the appropriate filters as noted in Table \ref{T:fotometri}. We noticed that only X1 and X7 have an association with nearby star groups or clusters. The CMDs for stars around {\it X1$\_1$} and {\it X7$\_1$} are displayed in Figure \ref{F:CMD1} and Figure \ref{F:CMD2} respectively.

\subsubsection{X-ray-to-optical flux ratios}

The ratios of X-ray to optical fluxes (F$_{X}$/F$_{V}$) for the optical counterparts can be used to distinguish between ULXs and AGNs \citep{2020MNRAS.499.2138A} (and references therein). Since we do not have simultaneous X-ray and optical observations, we calculated the ratio of F$_{X}$/F$_{V}$ using 2006 {\it Chandra} and 2005 {\it HST}/ACS (j95801020) data. Where F$_{X}$ is the flux observed in the range 2–10 keV and F$_{V}$ is the optical flux in the F550M filter. According to a study by \cite{2010MNRAS.401.2531A}, this ratio is defined in the range of 0.1-10 for AGNs. The calculated ratios for the optical counterparts of X1, X2, X5 and X7 are listed in Table \ref{T:tab6}.

\subsubsection{X-ray-Ultraviolet Spectral Correlations}

The distinct role played by the accretion disk in the emission process is well known for AGNs \citep{2010A&A...512A..34L, 2013A&A...550A..71V} and is well encapsulated in a X-ray-UV correlation cast in terms of the so-called optical spectral index $\alpha_{ox}$, as defined by \cite{1979ApJ...234L...9T}. The index is computed from the X-ray and UV flux densities at 2 keV and 2500\r{A} respectively. This relation was recently exploited by \cite{2019ApJ...873L..12S} to demonstrate an X-ray-UV correlation for a small sample of ULXs for which the optical counterparts are known. For comparison, we extracted $\alpha_{ox}$ for the candidate ULXs in NGC 1672. These $\alpha_{ox}$ values are listed in Table \ref{T:tab6}. 

\subsection{X-ray}

\subsubsection{Spectral fitting}

NGC 1672 was observed by {\it XMM-Newton} and {\it Chandra} in 2004 and 2006, respectively. In addition, it was observed 17 times by {\it Swift-XRT} between 2006 and 2020. The log of observations is given in Table \ref{T:observations}.

The {\it Chandra} ACIS (Advanced CCD Imaging Spectrometer) data were analyzed by using {\scshape ciao}\footnote{https://cxc.cfa.harvard.edu/ciao/} v4.12 software with its calibration package {\scshape caldb}\footnote{https://cxc.cfa.harvard.edu/caldb/} v4.9. The source and background photons were extracted with {\scshape specextract} task using circles with radii of 6$\arcsec$ and 12$\arcsec$, respectively.

The {\it XMM-Newton} data reductions were carried out using the {\scshape sas} (Science Analysis Software version 18.0). The {\it epchain} and {\it emchain} tasks were used to obtain EPIC-pn and MOS event files. The events corresponding to PATTERN$\leq$12 and PATTERN$\leq$4 with FLAG=0 were selected for EPIC-pn and MOS detectors, respectively. Source and background spectra of the candidate ULXs were extracted using the {\it evselect} task, with appropriate circular regions of radii 10$\arcsec$ $-$ 15$\arcsec$ and 20$\arcsec$ $-$ 30$\arcsec$, respectively. The background regions were selected from source free regions on the same chip containing the ULXs.
However, since the X9 source is localized on the EPIC-pn chip gap, we used only the MOS data for spectral fitting.
For the other ULXs, EPIC-pn and MOS spectra were fitted simultaneously in the 0.3$-$10 keV energy band and a constant scaling factor was taken in order to consider the cross calibration differences between the instruments. The files for the source and background spectra were obtained using the {\it dmgroup} task.

The {\it Swift-XRT} data sets were processed with HEASoft v6.29\footnote{https://heasarc.gsfc.nasa.gov/docs/software/heasoft/}, the tool {\it xrtpipeline} and calibration files CALDB v4.9. Data used were taken in Photon Counting (PC) mode in the range of 0.3–10 keV. The source and background photons were extracted circular regions of radii 18$\arcsec$ and 36$\arcsec$ respectively. Appropriate ancillary response files were generated with the task {\scshape xrtmkarf}. We combined the data taken in the same year (e.g. six datasets in 2012) for spectral fitting due to the short exposure time and low data statistics. \\

\cite{2011ApJ...734...33J} fitted the X-ray spectra with a {\it power-law} model and obtained
reasonable fits. In order to explore whether these fits could be improved, the spectra for the ULX candidates which have sufficient statistics were grouped at least 15 counts per energy bin.
We then fit the grouped spectra by using the
package, xspec v12.11 \citep{1996ASPC..101...17A} with the following models (in addition to the
{\it power-law}): several single component models such as the absorbed multi-color disk
blackbody ({\it diskbb}), {\it diskpbb} and {\it cutoffpl} respectively. We also fitted the spectra with the frequently used two-component models such as {\it power-law + diskbb} and {\it diskbb
+ comptt}.

\begin{figure}
\begin{center}
\includegraphics[width=\columnwidth]{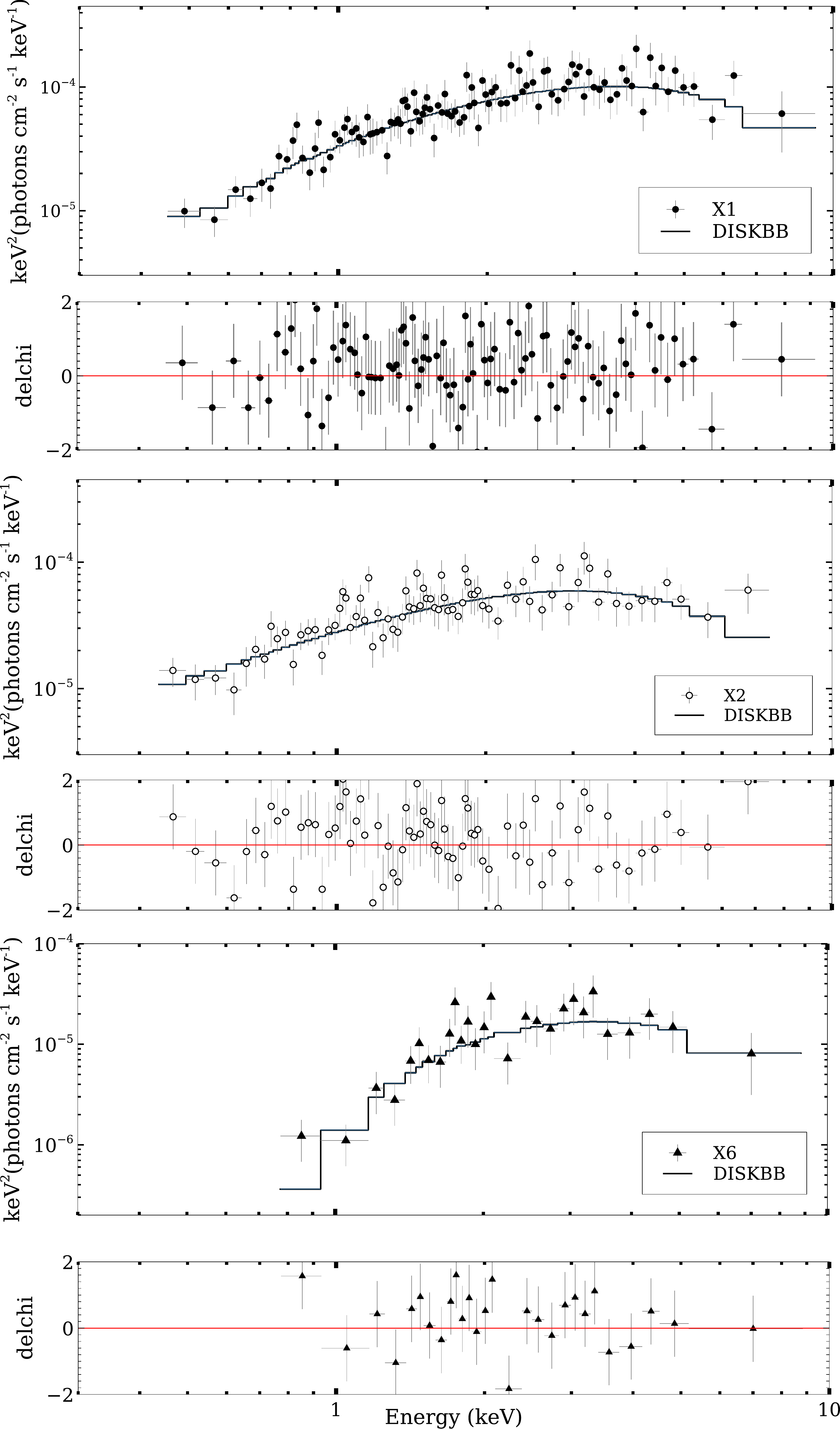}
\caption{Energy spectra of X1 (upper), X2 (center) and X6 (lower) by using {\it Chandra} data in the 0.3-10 keV energy range. The spectra were fitted with an {\it diskbb} model. The units of delchi is (data-model)/error.}
\label{F:xspec}
\end{center}
\end{figure}

In Table \ref{T:xmodel}, the well$-$fitted spectral model parameters for X1, X2 and X6 (deploying the {\it Chandra} data only) are given with unabsorbed flux values. The unabsorbed flux was calculated in the 0.3$-$10 keV energy band using the convolution model {\scshape cflux} available in {\scshape xspec} and corresponding luminosity was calculated using a distance of 16.3 Mpc \citep{2000AJ....119..612D}. The acceptable model-fits were selected according to null hypothesis probability. We consider all fits with P $>$ 0.05 corresponding to a confidence level >95\%.
We found the {\it diskbb} model fits to be statistically better than the other single component models at the 3$\sigma$ confidence level. The absorption component was left as a free parameter in the fitting procedure. Inclusion of two component models for these sources did not improve the fits significantly compared to {\it diskbb} model. The unfolded energy spectra of the best model-fits, for X1, X2 and X6, are shown in Figure \ref{F:xspec}.

The ULX candidates with low data statistics (i.e X5, X7, X8 and X9) were also fitted with the same single component models using C-statistics. However, none of the models gave reasonable results therefore the {\it power-law} model parameters are
 the accepted in our final calculations as given in the study of \cite{2011ApJ...734...33J}.
This is partially due to the lack of sufficient data for these sources leading to poor statistics. 
On the other hand, the {\it XMM-Newton} and {\it Chandra} data statistics are reasonable for X3 but a satisfactory fit was not achieved with one or two-component models. More complex combinations of models were not pursued as it further deteriorates the interpretation of the extracted parameters. We note that X4 is detected as a distinct source only in the {\it Chandra} observation. In the {\it XMM-Newton} and {\it Swift-XRT} observations, the source is not resolved from the galaxy center, and as such was not considered for further spectral analysis. \\

\subsubsection{Long-term and Short-term X-ray Variability} \label{sec:212}

We used all the available X-ray data to examine the long-term variability of the candidate ULXs. Fluxes of {\it XMM-Newton} and {\it Chandra} observations were obtained by using the {\it power-law} model in 0.3$-$10 keV energy band at 1-$\sigma$ confidence level. As the {\it Swift-XRT} observations do not have enough statistics, we combined the event files for each observation within a given year. This was applied to all the data taken in the years 2007, 2012, 2017 and 2019 respectively (see Table \ref{T:observations}). Unfortunately even the combined statistics proved insufficient thus in this case, we obtained the {\it Swift-XRT} fluxes from the respective count rates using {\it WebPIMMS}\footnote{
https://heasarc.gsfc.nasa.gov/cgi-bin/Tools/w3pimms/w3pimms.pl}. We used the {\it power-law} model parameter, $\Gamma$, and N$_{\mathrm{H}}$ obtained from the {\it Chandra} observation of each ULX. The source X5 was not detected in {\it XMM-Newton} and 2019 {\it Swift-XRT} so its flux values were calculated at the 3$\sigma$ upper limit. The light curves of the sources which have optical counterparts are shown plotted in Figure \ref{F:lcxrays}. The source X9 seems to exhibit more variability than the others.

We also performed a periodicity search for all of the candidate ULXs except X4. The background subtracted X-ray light curves of {\it Chandra} and {\it XMM-Newton} data were sampled at 0.1 s and 3.2 s, respectively, in the 0.3$-$10 keV energy band. The light curves were divided into six time intervals and the power spectrum density (PSD) of each interval was computed via a Fourier transform. 
The six PSDs were averaged to probe for temporal features. Our results showed no evidence for any significant periodicity $\geq$ 1.5$\sigma$ in any of the sources examined.

\subsubsection{Hardness–Intensity and Hardness-Ratio Diagrams}

\begin{figure}
\begin{center}
\includegraphics[width=\columnwidth]{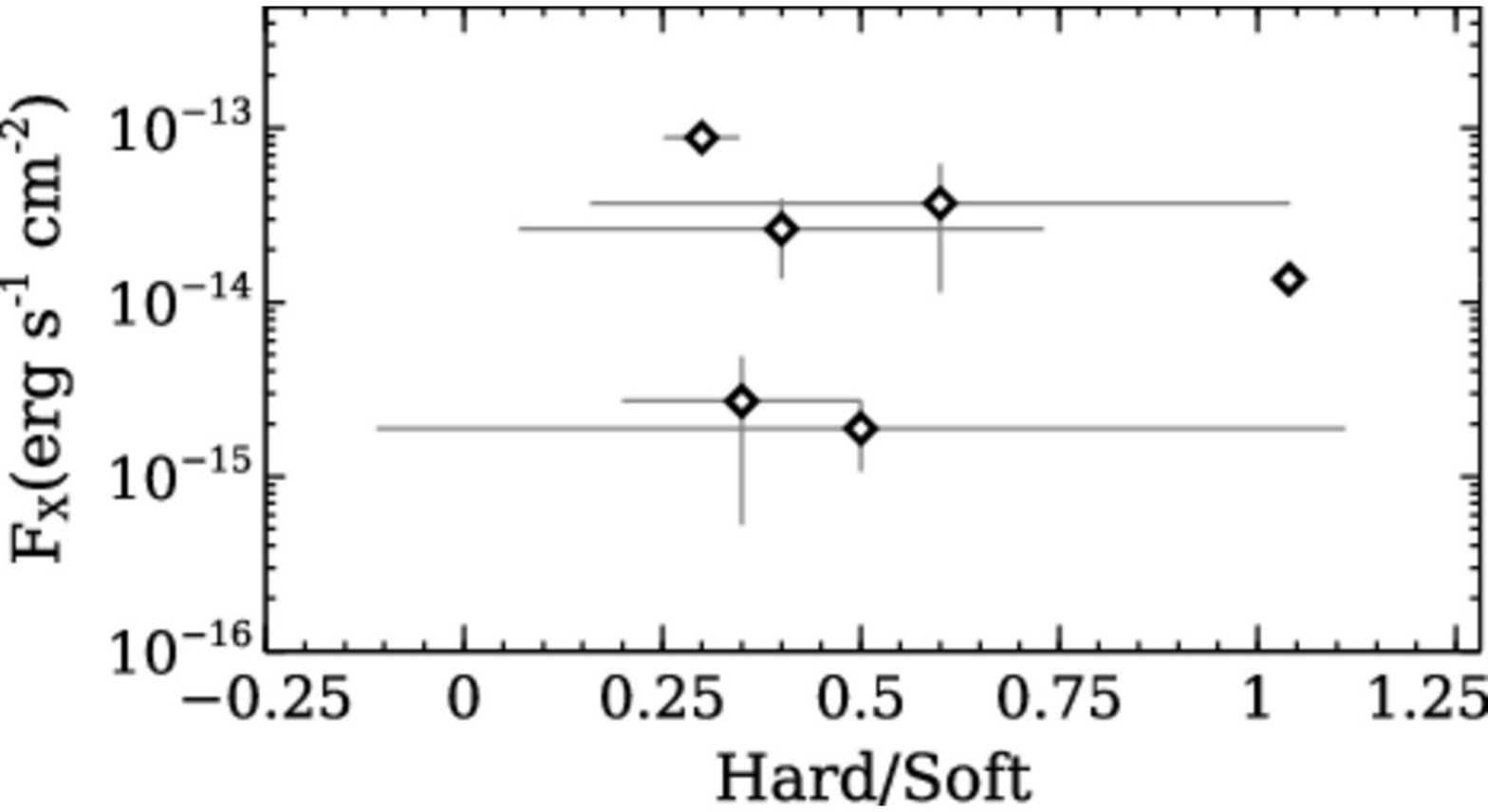}
\caption{The HID for X9 suggestive of a partial q-curve pattern.}
\label{F:HIDX9}
\end{center}
\end{figure}

In order to probe possible state transitions in the ULXs, we constructed hardness-intensity diagrams (HIDs) and hardness ratio diagrams (HRDs) for the sources that have optical counterparts. We included X9 in this group largely because of its X-ray variability although it has no identified optical counterpart. The observations used in determining the hardness rations (HRs) are listed in Table \ref{T:observations}. The HRs are calculated as the ratio of counts in the hard-to-soft energy bands where the respective bands are as follows: 0.3 - 2.0 keV (Soft) and 2.0 - 10.0 keV (Hard). For the HIDs, the fluxes specified in section \ref{sec:212} were used. The HIDs of the sources that have one optical counterpart are shown in the left panel of Figure \ref{F:HID}; the right panel of the figure depicts the HIDs of the sources that have two optical counterparts. The HID for X9 is displayed in Figure \ref{F:HIDX9}. The HRDs of these sources are displayed in Figure \ref{F:HRs}.

\section{Results and Discussion} \label{sec:3}
\subsection{Optical}

The problem with identifying the donor star in the optical emission of ULXs is that the flux contribution from the accretion disk, either directly or irradiated, is mostly unknown.

In the present study, we identified the possible optical counterparts of the ULXs in the galaxy NGC 1672 with precise astrometric calculations based on Chandra and HST data. Using HST UV/optical observations, we investigated in detail the optical variability, color and SED features of the optical counterparts to analyze their possible contribution to the observed optical emission.

Examining all the optical images listed in Table \ref{T:observations}, we identified a unique counterpart for X2 and X6, and two potential optical counterparts for X1, X5 and X7 respectively within the error radius of 0.$\arcsec$21 at 95\% confidence level.
It is noteworthy that X3 and X8, located in a spiral arm, and X4 located very close to the center of the galaxy, apparently have no optical counterparts or more likely, the counterparts are just not bright enough to be visible in the observations considered here. Unfortunately, X9 was not in the field of view of the {\it HST} archival observations and hence we are not in position to make any statement regarding its potential counterpart(s).

\begin{figure}
\begin{center}
\includegraphics[width=\columnwidth]{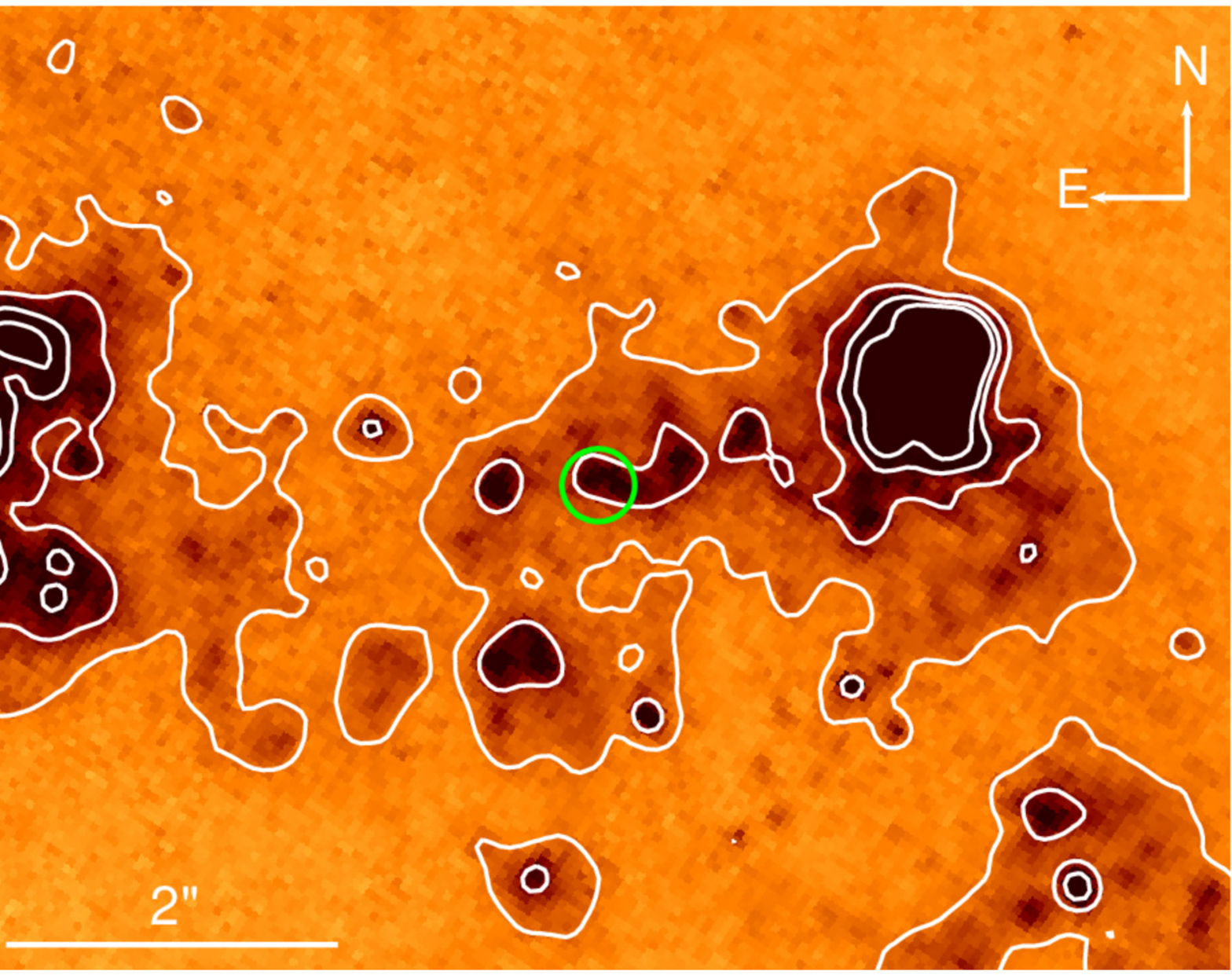}
\caption{{\it HST}/WFC F658N filter image showing the H${\alpha}$ Nebula around ULX X1. The position of X1 is shown within the astrometric error radius (green circle). White contour levels are plotted on the image.}
\label{F:halpha}
\end{center}
\end{figure}

The absolute magnitudes, (M$_{V}$), of the optical counterparts are listed in Table \ref{T:tab6}, and are seen to lie in the range $-$5 < M$_{V}$ < $-$7.5. This range is compatible with the values ($-$4 to $-$9) noted by \cite{2013ApJS..206...14G}. In addition, we also note that the observed counterparts appear to be faint in the V$-$band magnitude (m$_{V}$<23.5) similar to most of the known ULXs \citep{2011ApJ...737...81T, 2013ApJS..206...14G,2018ApJ...854..176V}. As a part of this study, we also examined the environment in the vicinity of the target ULXs to investigate if there were any associations between the ULXs and nearby star groups or clusters. Widely scattered star groups are located near X1 and X7 sources. Results of optical photometry indicate that these nearby bright stars have similar properties (such as colors and magnitudes) to the counterparts of X1 and X7. Also, when we examined the F658N image to search for existing star forming regions and check for possible ULX nebulae, we noticed that X1 is located in a HII region; the region is shown in Figure \ref{F:halpha}.

Of course, an intrinsic property of particular interest for any optical counterpart is its mass.
The mass of the donor (M$_{d}$) (in an accreting system for example) is useful in not only categorizing the binary system (i.e., a low-mass X-ray binary (LMXB) or a high-mass X-ray binary (HMXB) etc) but is fundamentally important in determining many of the emission processes associated with the binary system. 
Although a consistent mass range of the donor is not provided in the literature, we adopt the mass scale given in the recent studies of \cite{2019ApJ...871..122J,2020ApJ...890..150C,2021ApJ...912...31H}, i.e., M$_{d}$ >8 M$_{\odot}$ for a HMXB; 3 M$_{\odot}$< M$_{d}$ <8 M$_{\odot}$ for an intermediate X-ray binary, and M$_{d}$<3 M$_{\odot}$ for a LMXB. 
When optical emission is observed in the positional uncertainty of an ULX, it is potentially an optical counterpart. The optical counterparts of the ULXs given in Table \ref{T:fotometri} are likely donor stars of HMXBs since they have been detected at least in the one of {\it HST} filters.
For X3, X4 and X8, we were not able to identify any optical counterparts, probably because they are too faint; see studies by \cite{2021ApJ...912...31H, 2021ApJ...922..178R}, and references therein, who suggest an {\it HST} observation threshold for stars with a mass <3 M$_{\odot}$ at a distance of $\sim$10 Mpc.

As a check to determine whether any of the target sources are possibly background AGNs, we determined the F$_{X}$/F$_{V}$ ratios. These ratios range from $\sim$ 11 to $\sim$ 50, which confirms that these sources are unlikely to be AGNs for which the ratio falls in a significantly lower range i.e., (0.1< (F$_{X}$/F$_{V}$) <10, \cite{2010MNRAS.401.2531A}). In addition, we extracted the spectral optical index \citep{1979ApJ...234L...9T}, $\alpha_{ox}$, which has been shown \citep{2019ApJ...873L..12S} to exhibit a distinctly different correlation for ULXs as compared to AGN. The extracted values of $\alpha_{ox}$ are listed in Table \ref{T:tab6} and they agree well with the typical values found for other well-studied ULXs.

\begin{figure*}
\begin{center}
\includegraphics[angle=0,scale=0.30]{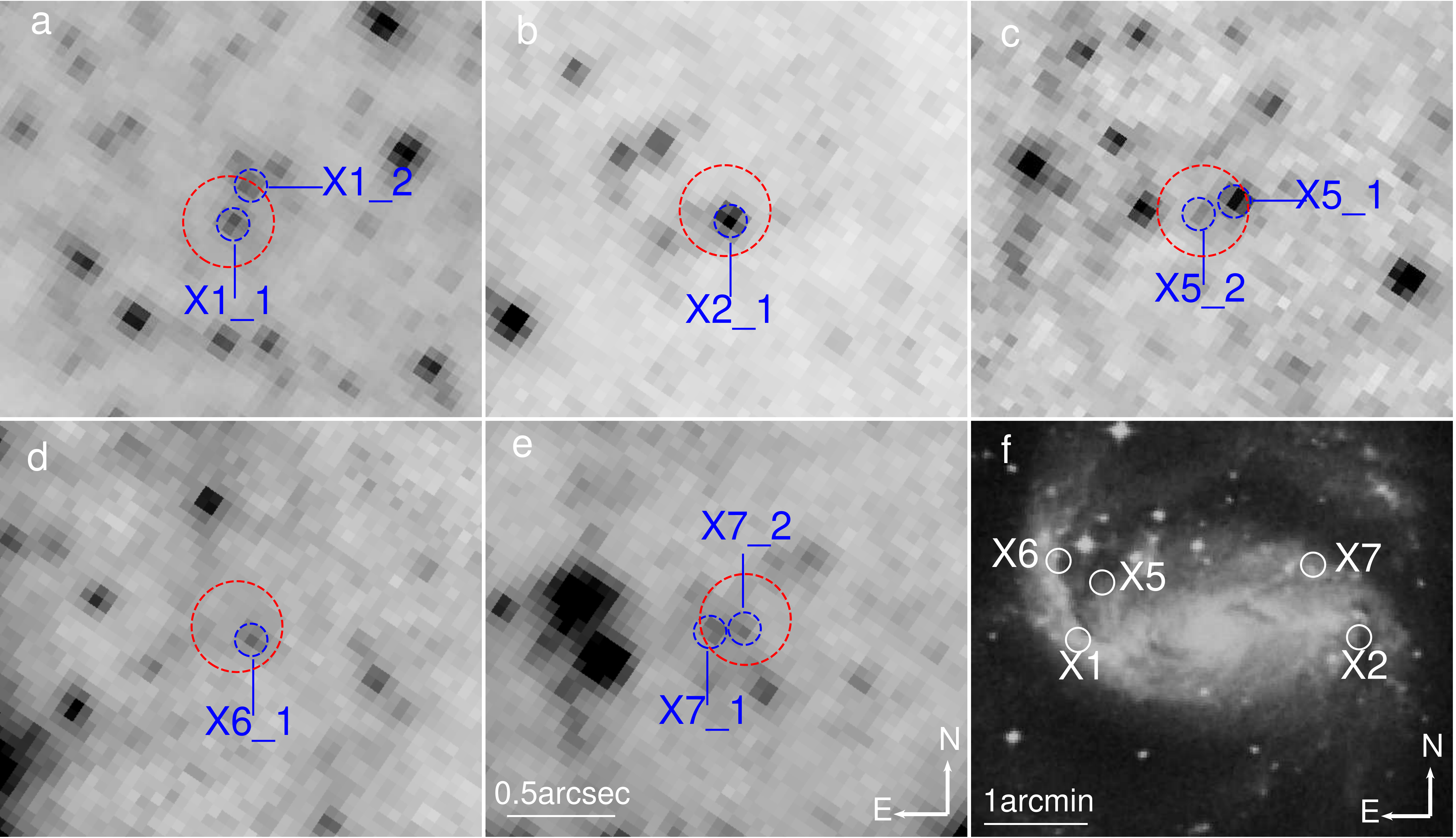}
\caption{The identified optical counterparts of the five ULXs are shown on the F814W images. Red dashed circles indicate the astrometric error radius of the optical counterparts given in Table \ref{T:astrometry}. Blue dashed circles represent the center coordinates of each optical counterpart. North is up and east is left in all panels. Panels (a-e) are the same scale. In panel f shows positions of ULXs which have optical counterparts on the DSS image.}
\label{F:can}
\end{center}
\end{figure*}

In the remaining part of this section, we mention some of the salient features of the optical counterparts identified for each of the ULX candidates:

{\it X2} and {\it X6}: A unique counterpart was identified for both X1 and X6 within the 0.$\arcsec$21 astrometric error radius. The optical counterparts are shown in the F814W image in Figure \ref{F:can}. Dereddened magnitudes and flux values of the counterparts in each filter are given in Table \ref{T:fotometri}.
{\it X2$\_1$} appears to be a faint source in the V$-$band and is relatively bright in the UV band. Interestingly {\it X6$\_1$} was detected only in the I-band with 25.33 $\pm$ 0.11 mag.

The intrinsic colors of {\it X2$\_1$} are as follows: (B-V)$_{0}$ $=$ 0.03 and (V-I)$_{0}$ $=$ 0.05. Using the Schmidt–Kaler table, the spectral class of {\it X2$\_1$} was determined to be B8$-$A3, a supergiant donor star. We also tried to confirm the spectral class of {\it X2$\_1$} using the reddening corrected SED as given in Figure \ref{F:SEDsX2}. The SED for the source is adequately fitted (a $\chi^2_{\nu}$ value of 0.93 (dof=5) at 95\% confidence level) with a {\it blackbody} model and a temperature of 10$^{4}$ $\pm$ 113 K. In this case assuming the majority of the emission arises from the donor star, the temperature and the luminosity of {\it X2$\_1$} correspond to a B7 type supergiant according to the table given in \cite{1981Ap&SS..80..353S}. We also determined the same spectral type of {\it X2$\_1$} based on the CK04 standard stellar spectra templates. On the other hand, the blackbody emission could also originate from irradiation of the accretion disk.Therefore, both X-ray and optical variability need to be investigated. While X2 is quite variable in the X-ray range (factor of $\sim 8$), it is difficult to comment on the optical variability considering we have only two observations 14 years apart, F550M in 2005 and F555W in 2019. Even in these two observations no optical variability was seen.

\begin{figure}
\begin{center}
\includegraphics[width=\columnwidth]{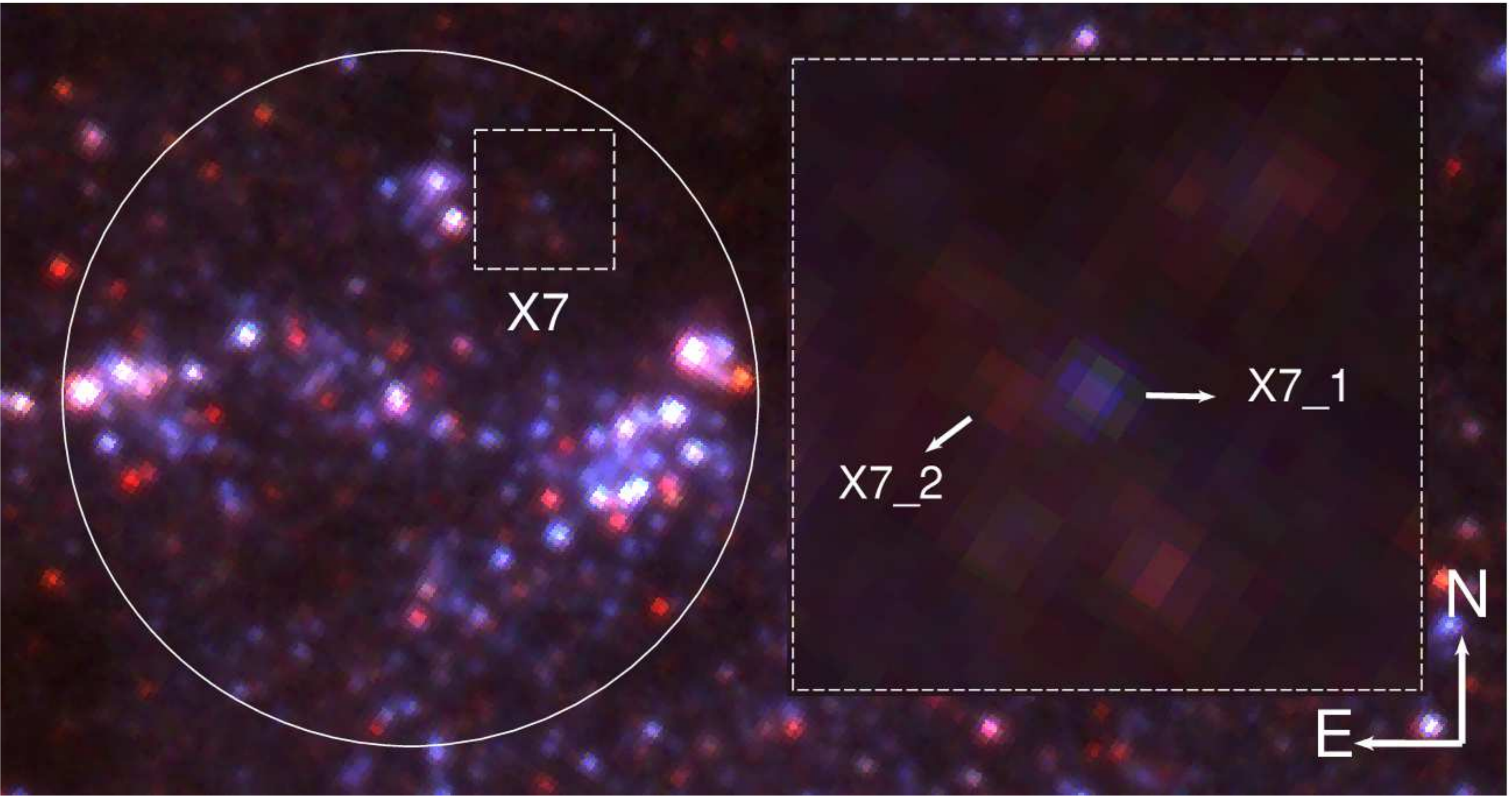}
\includegraphics[width=\columnwidth]{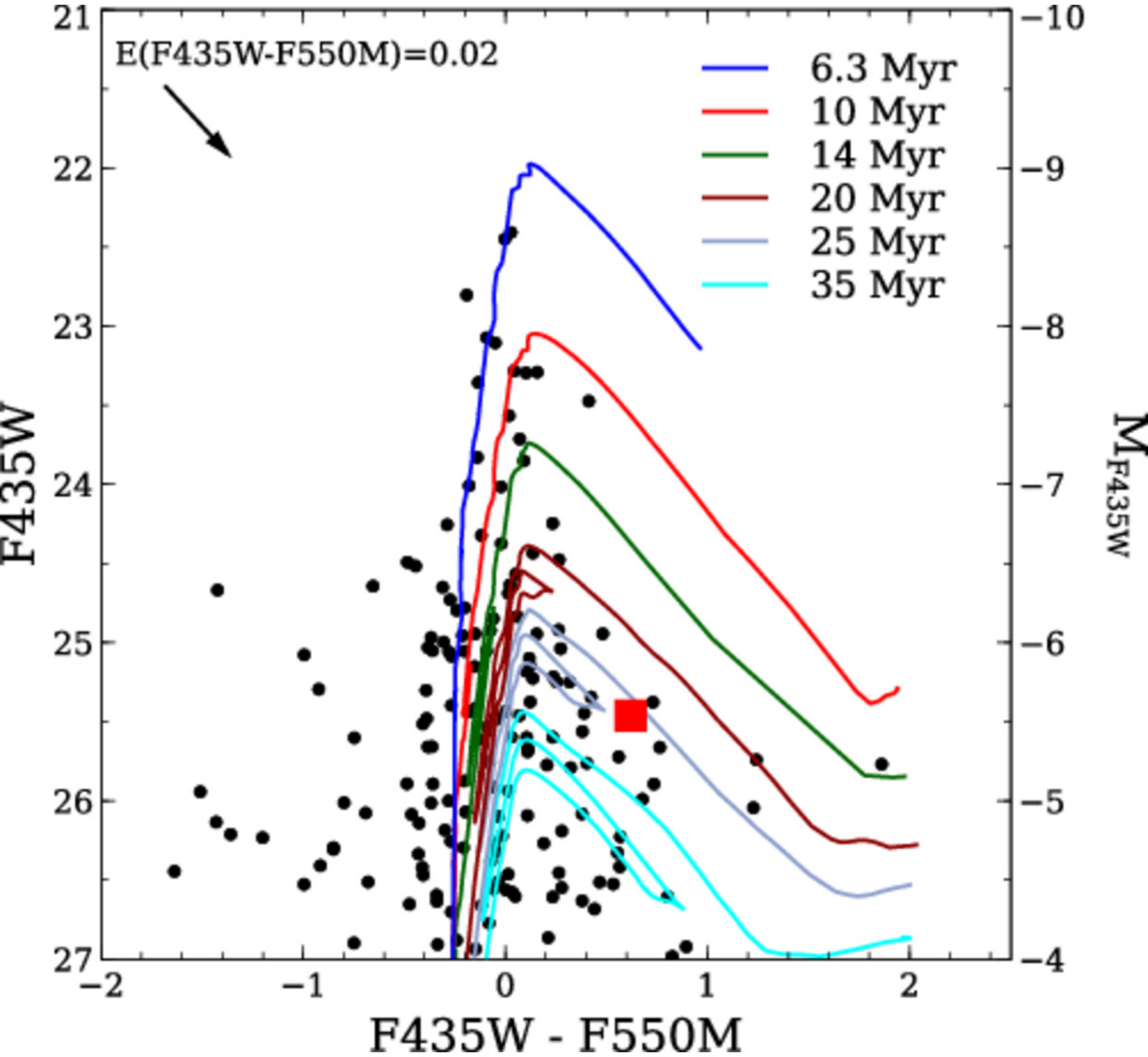}
\caption{Upper: Optical counterparts of X7 and their environments are shown on the {\it HST} RGB image. A large group of stars is indicated by a white circle with 2.$\arcsec$5 radius. Zoomed view of the optical counterparts for clarity is given upper right. Lower: CMD for the {\it X7$\_1$}, a large group of stars, and field stars around {\it X7$\_1$}. The red square and black dots represent the {\it X7$\_1$} and field stars within the given region. The isochrones have been corrected for extinction of A$_{V}$ $=$0.065 mag and the black arrow shows the reddening line.}
\label{F:CMD2}
\end{center}
\end{figure}

{\it X1, X5 and X7}: Two possible optical counterparts are identified for each of these ULX candidates; see Figure \ref{F:can}. The calculated dereddened magnitudes and fluxes of the counterparts are given in Table \ref{T:fotometri}. A careful examination of all the {\it HST} images shows the counterparts {\it X1$\_1$}, {\it X5$\_1$} and {\it X7$\_1$} to be brighter in the UV (m$_{F275W, F336W}>22$) compared to the other filters. On the other hand {\it X1$\_2$} is only visible in the I-band filter (m$_{I} = 24.22$ mag). It is possible that this source is similar in nature to {\it X6$\_1$}. We note that {\it X5$\_1$} is detected in all images taken between 2005 to 2019 while {\it X5$\_2$} is only seen in the 2019 UVIS images. This suggests that {\it X5$\_1$} is a persistent source whilst {\it X5$\_2$} is a variable source. As noted in Table \ref{T:fotometri}, {\it X7$\_2$} is detected in three ACS filters. This source appears reddish in color in the {\it HST} true color image depicted in Figure \ref{F:CMD2}. It is possible this source is a red supergiant.\\
\\
The CMD, deploying the filters F435W versus (F435W-F550M), was used to investigate the association of optical counterparts to nearby stars or group of stars. For this case, {\it X1$\_1$} and {\it X7$\_1$} are the best candidates, as seen in the top panels of Figures \ref{F:CMD1} and \ref{F:CMD2}. If we consider the age intervals of the star groups close to the counterparts, then {\it X1$\_1$} and {\it X7$\_1$} can be associated with stars with ages (10-13) Myr and (25-30) Myr, respectively. We also derived color indices for the counterparts (see Table \ref{T:tab6}) then using intrinsic colors and the absolute magnitudes of the Schmidt–Kaler table \citep{1982lbg6.conf.....A}, the spectral types were determined to be early type B$-$A supergiants.

Optical counterparts {\it X1$\_1$} and {\it X1$\_2$} are located within a HII region as seen in the F658N image (see Fig. \ref{F:halpha}). Source {\it X1$\_2$} is not resolved in the F658N band and it is only detected in the F814W filter. Using a circular region with a radius of 0.$\arcsec$21 (and following the procedure given in section \ref{sec:sed}), we calculated the total F658N flux as 1.01$\times$10$^{-18}$ erg cm$^{-2}$ s$^{-1}$ \AA$^{-1}$. This calculated F658N flux was determined to be $\sim$ 40\% of the total flux (F658N$+$continuum background).

The SED for the optical counterparts {\it X1$\_1$} is adequately fitted with {\it power-law}, F$\propto$ $\lambda^{\alpha}$, with a photon index, $\alpha$$=$$-$ 2.13 $\pm$ 0.16 at 95\% confidence level. The $\chi^2_{\nu}$ is found as 0.87 with 4 dof. The reddening corrected SED of {\it X1$\_1$} is shown in Figure \ref{F:SEDsX1}. If the optical emission originates mostly from an accretion disk then the optical SED should have a roughly {\it power-law} form with a photon index of $\alpha$ about -2.3 and significant optical variability \citep{2011ApJ...737...81T}. The extracted $\alpha$ is consistent with the expected value within errors. Thus, for this source, the optical emission may be dominated by an intrinsic emission from an accretion disk. The m$_{F336W}$ $\sim$ 0.6 mag. variability detected in the UVIS/F336W images for this source, over an interval of about 1.5 years, further supports this scenario.

\subsection{X-ray}

The values given in Table \ref{T:xmodel} show the well$-$fitted spectral model parameters for X1, X2, and X6. The unfolded energy spectra of these sources obtained via the use of the {\it diskbb} model are shown in Figure \ref{F:xspec}. These sources tend to exhibit relatively high values for the inner disk temperature (i.e., $T_{in}>$ 1.2 keV). This is a known feature of ULX spectra. 

\begin{figure}
\begin{center}
\includegraphics[width=\columnwidth]{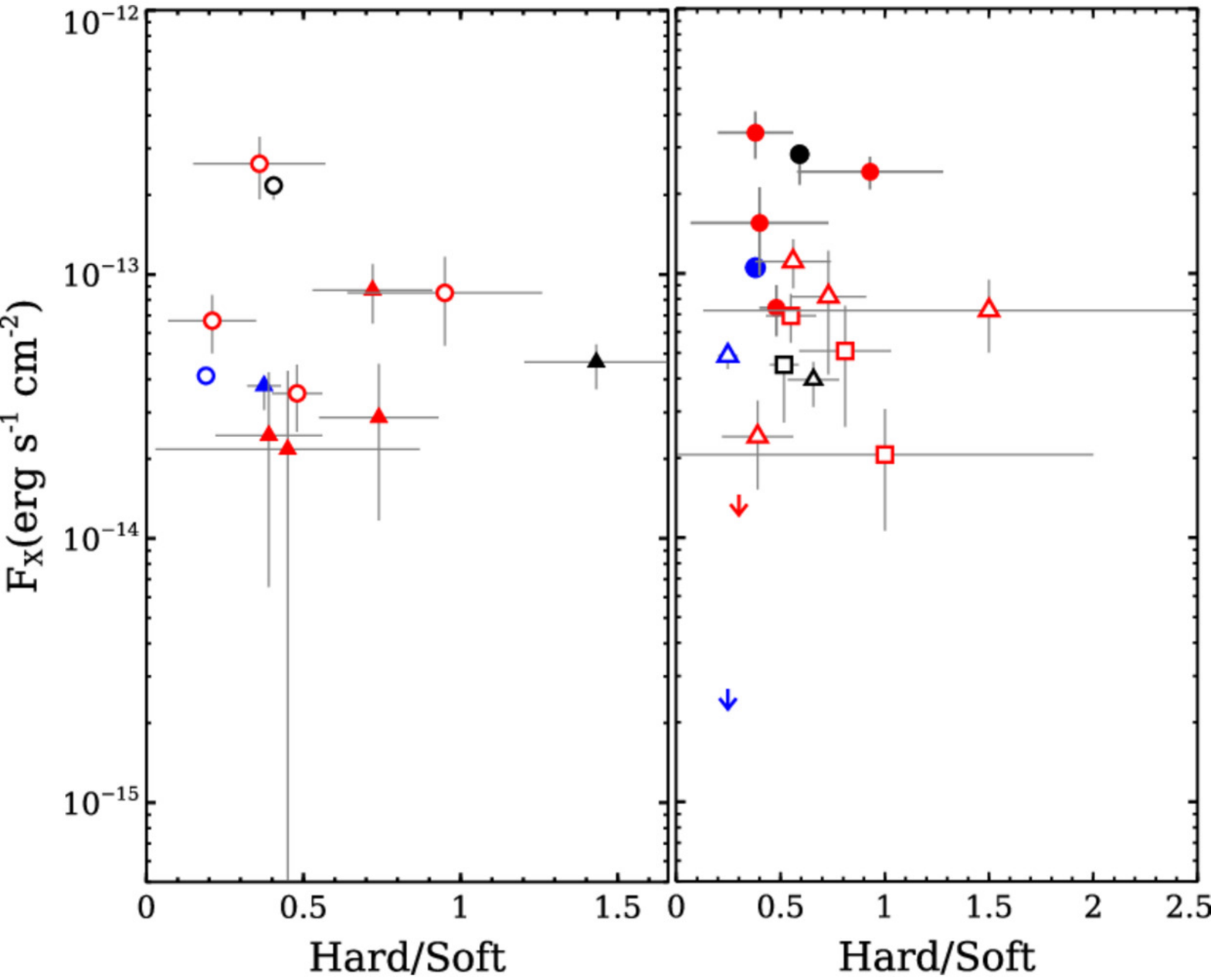}
\caption{Left panel: The hardness-intensity diagram (HID) for X2 (empty circles) and X6 (filled triangles). Right panel: The HID for X1 (filled circles), X5 (empty squares) and X7 (empty triangles). The downward arrows indicate the 3$\sigma$ upper limits where the X5 source cannot be detected. In both panels, black, blue and red colors represent {\it Chandra}, {\it XMM-Newton} and {\it Swift-XRT} observations, respectively.}
\label{F:HID}
\end{center}
\end{figure}

We revisit X9 when we consider the long-term variability of the ULXs. Nonetheless, the fits for the other sources are in reasonable (statistical) agreement with those found by \cite{2011ApJ...734...33J} but given the limited statistics, we refrain from drawing firm conclusions regarding the choice of spectral models and the extracted parameters.\\
\\
The HID is a proven diagnostic tool in tracking spectral transitions in the case of BH binaries. These transitions typically trace out a q-shape contour in the HID as the binary transitions from the low-hard state through the hard-intermediate, soft-intermediate, and finally back down to the low-hard state, and in some cases, into the lowest intensity state corresponding to the quiescent state (e.g., \citealp{2003A&A...411..553B,2003MNRAS.342.1041D,2006ARA&A..44...49R,2010LNP...794...53B}). We constructed HIDs using the hardness ratios determined from the count rates in the soft (0.3 - 2.0 keV) and hard states (2.0 - 10.0 keV) respectively, in order to probe similar transitions in the ULX candidates \citep{2012ApJ...750..152S}. The HIDs of X2 (empty circles), X6 (filled triangles) are given in the {\it left} panel of Figure \ref{F:HID}, and the HIDs of X1 (filled circles), X5 (empty squares) and X7 (empty triangles) are displayed in the {\it right} panel of Figure \ref{F:HID}. Furthermore, the HID of X9 is given in Figure \ref{F:HIDX9}. Although a complete q-curve shape is not discernible, X9 does, however, hint at a partial loop in the HID plane, thus suggesting possible transitions between spectral states.

\begin{figure}
\begin{center}
\includegraphics[width=\columnwidth]{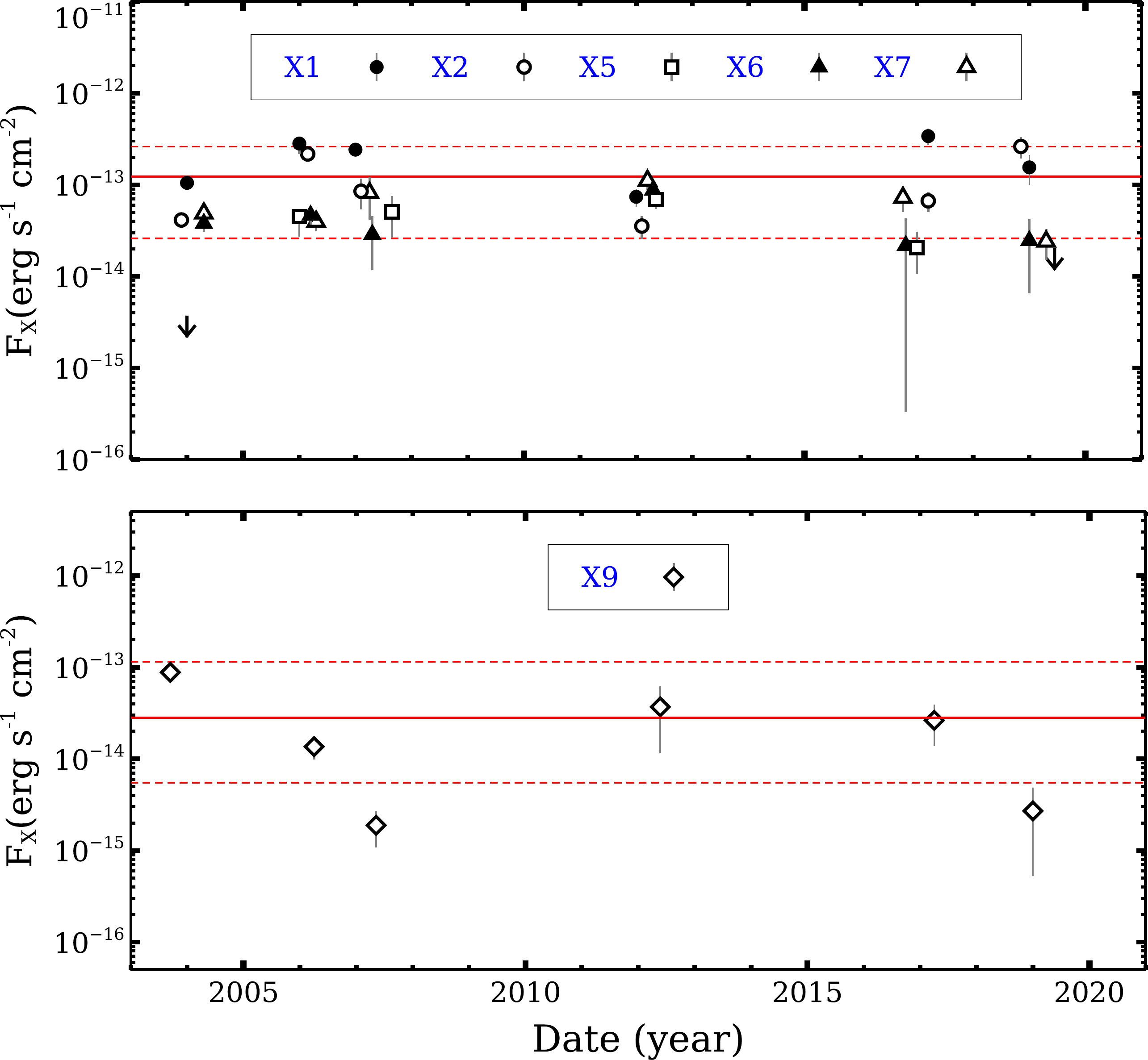}
\caption{Upper: The long-term X-ray light curve for the ULX candidates which have optical counterparts. The vertical arrows represent 3$\sigma$ upper limits when the source is not detected. Lower: The long-term X-ray light curve of X9. In both panels the solid line indicates the mean value of the X-ray flux; a measure of the variability is provided by the $\pm$ 3$\sigma$ levels shown as dotted lines. Observations are from {\it XMM-Newton} (2004), {\it Chandra} (2006) and {\it Swift-XRT} (2007; 2012;2017 and 2019).}
\label{F:lcxrays}
\end{center}
\end{figure}

\begin{figure}
\begin{center}
\includegraphics[width=\columnwidth]{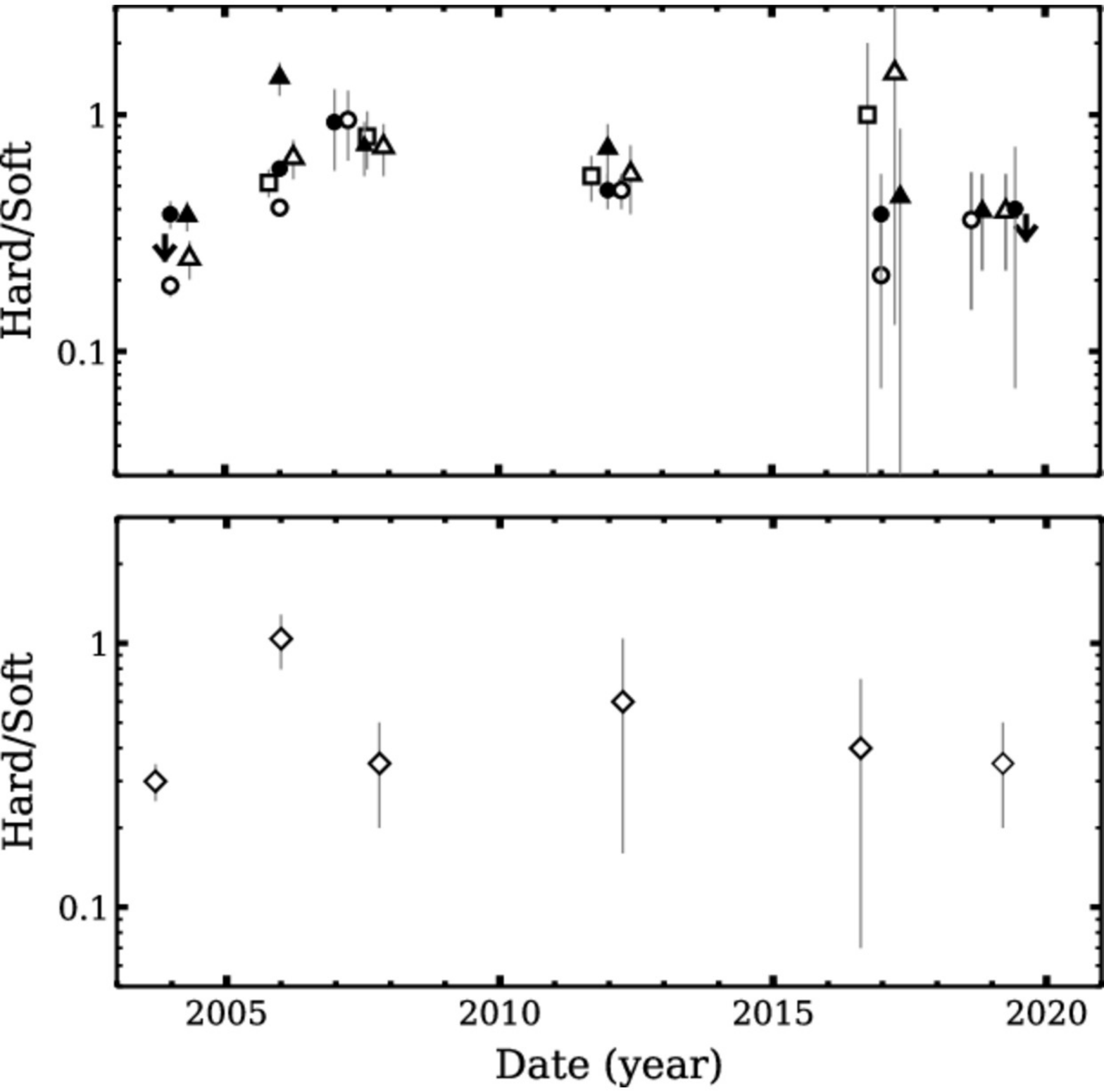}
\caption{The long-term variation of the hardness ratios of X1, X2, X5, X6 and X7 (upper) and X9 (lower). The downward arrows indicate the 3$\sigma$ upper limits where the X5 source cannot be detected. Observations are from {\it XMM-Newton} (2004), {\it Chandra} (2006) and {\it Swift-XRT} (2007; 2012;2017 and 2019). The same symbols given in Figure \ref{F:lcxrays} are used.}
\label{F:HRs}
\end{center}
\end{figure}

We explore this variability more systematically by constructing long-term light curves for the selected ULXs using all available {\it Swift-XRT}, {\it XMM-Newton}, and {\it Chandra} observations. The light curves of the sources that have optical counterparts are displayed in Figure \ref{F:lcxrays}. In each case, the solid line indicates the mean value of the X-ray flux; a measure of the variability is provided by the $\pm$ 3$\sigma$ levels shown as dotted lines.
As is clear from the Figure \ref{F:lcxrays}, the flux for X9 varies almost a factor 50 $\pm$ 18 between the different observations. In addition to the X9, the transient source X5 also exhibits high degree variability (almost a factor of 30) in the long-term. It should be noted that while the source was undetected in the 2004 {\it XMM-Newton} and 2019 {\it Swift-XRT} observations, it has the highest flux from the longest exposure in 2012 {\it Swift-XRT} data. The occurrence of high flux variability in ULXs may provide some evidence for exhibiting bi-modal flux distributions as might be expected for sources related to propeller transitions. As discussed in several studies, probing some ULX transient systems with such a bi-modal flux distribution may be a way to identify a ULX hosts a NS without detecting pulsations \citep{2018MNRAS.476.4272E,2020MNRAS.491.1260S,2022MNRAS.510.4355A}. 
However, we note that the statistics for the X5 and X9 sources are very low and the bi-modal nature is not evident in all the datasets we used.

 \citealp{2020MNRAS.491.1260S} e.g. SRC 279969) for X5 and X9 these distributions consistent with bi-modality.
We calculated the variability factor as the ratio of maximum to minimum fluxes; this factor is found to be 5 $\pm$ 0.4, 8 $\pm$ 2.9 and 5 $\pm$ 1.9 for X1, X2 and X7 respectively. For the remaining sources, i.e., X3, X6 and X8, the factor is $\leq$4. From this it is readily transparent that sources X5 and X9 show significant variability and thus are potential candidates for additional long-term monitoring. The corresponding long-term variation of the hardness ratio (HR) of X1, X2, X5, X6, X7 and X9 is displayed in Figure \ref{F:HRs}. The HR too shows variability among the sources but is not as striking as that exhibited by the flux.

Finally, we note that we extracted the optical spectral index $\alpha_{ox}$ for the candidate ULXs that have optical counterparts. The values are listed in Table \ref{T:tab6}, and are in good agreement with those reported by \cite{2019ApJ...873L..12S}. Importantly, these indices are quite different from those reported for AGN (see \cite{2010A&A...512A..34L, 2013A&A...550A..71V}) and thus this distinction provides another possible criterion for distinguishing ULXs from AGN.

\section{Summary and Conclusions} \label{sec:4}

A recent multiwavelength study of NGC 1672 identified multiple bright X-ray sources including 9 ULX candidates. In this study, we performed an optical, spectral, and a temporal analysis of those 9 ULX candidates. We deployed archival optical data from {\it HST}, and X-ray data from {\it Chandra}, {\it XMM-Newton}, and {\it Swift-XRT}. Specifically, our study focuses on using the precise source positions obtained via improved astrometry based on the {\it Chandra} and {\it HST} observations to search for and identify potential optical counterparts for the ULX candidates. We summarize our main findings as follows:\\

\begin{enumerate}
\item We identified unique optical counterparts for X2 and X6.\\ 

\item In the case of X1, X5 and X7, we were able to isolate two optical counterparts for each source.\\ 
\item No optical counterparts were found for X3, X8 and X4 within the respective error circles. The source X9, unfortunately did not have any {\it HST} data, so it could not be investigated further.\\

\item Photometric results of bright star groups in the vicinity of X1 and X7 suggest that their counterparts have similar properties to the nearby stars.\\

\item We constructed CMDs to estimate the age of the optical counterparts ({\it X1$\_1$} and {\it X7$\_1$}). The extracted age values of {\it X1$\_1$} are the order of $\leq 20$ Myr, whereas {\it X7$\_1$} is somewhat older i.e., 25-30 Myr.\\

\item The absolute magnitudes ($-$5 <M$_{V}$< $-$7.5) and spectral types (B-A) of the identified optical counterparts in NGC 1672 are compatible with optical companions of ULXs in other galaxies; see \cite{2013ApJS..206...14G} and \citealp{2011ApJ...737...81T}. Moreover, the optical counterparts appear to be faint in the V$-$band and relatively brighter in the UV band.\\

\item The SED for the counterpart {\it X1$\_1$} is well-fitted with {\it power-law}, F$\propto$ $\lambda^{\alpha}$, with a photon index, $\alpha$$=$$-$ 2.13 and also shows 0.6 mag. variability in UV. These findings are consistent with optical emission arising primarily from an accretion disk. Furthermore, the SED of {\it X2$\_1$} is adequately fitted with a {\it blackbody} model and a temperature of 10$^{4}$ K. In this case, the emission could come from the donor star and/or result from irradiation of the disk. \\

\item The sources X5 and X9 exhibit high X-ray variability, with the flux varying by factor of $\sim$ 30 and 50 times, respectively, over observations spanning a time period of 2004 to 2019. Due to the sparse coverage of the data, it is difficult to interpret that the high variability exhibits bi-modal flux distribution that could indicate the propeller effect. This level of variability together with the fact that X9 displays a partial q-curve track in the hardness-intensity diagram (possibly indicating spectral transitions) makes it an interesting source for further investigation.

\end{enumerate}

\begin{table*}
\centering
\caption{The log of optical and X-ray observations.}
\begin{tabular}{cccccc}
\hline
Instrument & ObsID & Date & Exp & Filter \\
& & (YYYY-MM-DD) & (ks)&\\
\hline
\multicolumn{5}{c}{Optical observations} \\
\\
ACS/WFC & j95801010$^{a}$ & 2005-08-01 & 2.44 & F435W \\
ACS/WFC & j95801020$^{a}$ & 2005-08-01 & 2.44 & F550M \\
ACS/WFC & j95801040$^{a}$ & 2005-08-01 & 2.44 & F658N \\
ACS/WFC & j95801030$^{a}$ & 2005-08-01 & 2.44 & F814W \\
WFC3/UVIS & idxr12040$^{b}$ & 2019-06-24 & 2.87 & F275W \\
WFC3/UVIS & idxr12030$^{b}$ & 2019-06-24 & 2.48 & F336W \\
WFC3/UVIS & idxr12050$^{b}$ & 2019-06-24 & 1.50 & F555W \\
WFC3/UVIS & idxr13040$^{c}$ & 2019-06-25 & 2.87 & F275W \\
WFC3/UVIS & idxr13030$^{c}$ & 2019-06-25 & 2.48 & F336W \\
WFC3/UVIS & idxr12050$^{c}$ & 2019-06-26 & 1.50 & F555W \\
WFC3/UVIS & ieb336020$^{b}$ & 2020-11-27 & 0.71 & F336W \\\\

\multicolumn{5}{c}{X-ray observations} \\
\\
{\it XMM-Newton} & 0203880101 & 2004-11-27 & 49.91& \\
{\it Chandra} & 5932 & 2006-04-30 & 40.00& \\
Swift-XRT&35882001 &2006-12-19 &1.84&\\
Swift-XRT&35882002 &2007-03-18 &2.43&\\
Swift-XRT&35882003 &2007-10-31 &1.01&\\
Swift-XRT&46271001 &2012-01-14 &0.77&\\
Swift-XRT&46271002 &2012-09-06 &1.11&\\
Swift-XRT&46271003 &2012-09-08 &0.49&\\
Swift-XRT&46271004 &2012-09-13 &0.55&\\
Swift-XRT&46271005 &2012-10-07 &8.87&\\
Swift-XRT&46271006 &2012-10-12 &7.56&\\
Swift-XRT&35882005 &2017-08-16 &1.99&\\
Swift-XRT&35882006 &2017-08-17 &1.85&\\
Swift-XRT&35882008 &2017-08-20 &2.23&\\
Swift-XRT&35882009 &2017-08-21 &2.96&\\
Swift-XRT&95188001 &2019-07-16 &0.65&\\
Swift-XRT&95188002 &2019-07-22 &0.88&\\
Swift-XRT&95188004 &2019-12-04 &0.69&\\
\hline
\end{tabular}
\\$^{a}$ These observations cover the eight ULXs (X1-X8), $^{b}$ X1, X5 and X6 $^{c}$ X2, X3, X4, X7 and X8 positions. \\
\label{T:observations}
\end{table*}

\begin{table*}
\centering
\caption{Coordinates of the X-ray/optical reference sources and ULXs.}
\begin{tabular}{cccrccccccc}
\hline
\multicolumn{7}{c}{{\it {\it Chandra}} ACIS-S X-ray sources (ObsID 5932) identified in {\it HST} observation (j95801030)}\\
\hline
Source number & {\it Chandra} R.A.& {\it Chandra} Dec.& Net Counts$^{a}$ & {\it HST} R.A.& {\it HST} Dec.& Offset$^{b}$ \\
... & (hh:mm:ss.sss) & ($\degr$ : $\arcmin$ : $\arcsec$) & & (hh:mm:ss.sss) & ($\degr$ : $\arcmin$ : $\arcsec$) & ($\arcsec$)\\
\hline
Ref.1 & 4:45:53.305 & -59:15:28.02 & 70.65 $\pm$ 8.54 & 4:45:53.313 & -59:15:27.95 & 0.17 \\
Ref.2 & 4:45:44.536 & -59:15:35.21 & 37.62 $\pm$ 6.25 & 4:45:44.541 & -59:15:35.14 & 0.13 \\
Ref.3 & 4:45:36.973 & -59:16:54.91 & 13.87 $\pm$ 3.87 & 4:45:36.989 & -59:16:54.76 & 0.35 \\
Ref.4 & 4:45:29.833 & -59:15:07.04 & 62.54 $\pm$ 8.06 & 4:45:29.847 & -59:15:07.05 & 0.20 \\
\\
\multicolumn{6}{c}{{\it Chandra} and corrected optical coordinates of optical counterparts} & Position uncertainty (${\arcsec}$)$^{c}$\\
\\
X1 & 4:45:52.823 & -59:14:56.15 & 1305.25 $\pm$ 36.51 & 4:45:52.834 & -59:14:56.08 & 0.21 \\
X2 & 4:45:31.603 & -59:14:54.68 & 955.95 $\pm$ 31.26 & 4:45:31.612 & -59:14:54.59 & 0.21 \\
X3 & 4:45:33.965 & -59:14:42.03 & 771.71 $\pm$ 28.07 & 4:45:33.975 & -59:14:41.93 & 0.21 \\
X4 & 4:45:42.166 & -59:14:52.17 & 480.53 $\pm$ 29.91 & 4:45:42.175 & -59:14:52.08 & 0.21 \\
X5 & 4:45:50.996 & -59:14:22.96 & 214.59 $\pm$ 14.83 & 4:45:51.006 & -59:14:22.87 & 0.21 \\
X6 & 4:45:54.289 & -59:14:10.44 & 144.52 $\pm$ 12.29 & 4:45:54.299 & -59:14:10.35 & 0.21 \\
X7 & 4:45:35.078 & -59:14:12.68 & 125.82 $\pm$ 11.40 & 4:45:35.087 & -59:14:12.59 & 0.21 \\
X8 & 4:45:28.453 & -59:14:33.39 & 105.65 $\pm$ 10.39 & 4:45:28.463 & -59:14:33.30 & 0.21 \\
X9$^{d}$ & 4:45:34.317 & -59:12:55.95 & 61.70 $\pm$ 8.00 & ... & ... & ... \\
\hline
\label{T:astrometry}
\end{tabular}
\\
Notes: $^{a}$The {\it Chandra} counts are given in the 0.3-10 keV energy range with 1-$\sigma$ errors using {\scshape xspec}.
$^{b}$Offsets are given at 95\% confidence level of the {\it Chandra}/{\it HST} reference sources.
$^{c}$Position uncertainties (error radius) are given at 95\% confidence level.
$^{d}$X9 was not observed by {\it HST}.
\end{table*}

\begin{table*}
\centering
\caption{The calculated dereddened Vega magnitudes and fluxes of the possible optical counterparts.}
\begin{tabular}{cccccccccc}
\hline
Optical counterparts & UVIS/F275W & UVIS/F336W & UVIS/F555W & ACS/F435W & ACS/F550M & ACS/F658N & ACS/F814W \\
\hline
\multicolumn{7}{c}{Vega Magnitude} \\
\hline
{\it X1$\_1$} & 22.37 $\pm $0.05 & 22.63 $\pm$ 0.04 & 24.16 $\pm $0.04 & 23.90 $\pm$ 0.03 & 23.93 $\pm$ 0.05 & 23.12$\pm$ 0.44 & 24.01$\pm$ 0.17\\
{\it X1$\_2$} & ... & ... & ... & ... & ... & ... & 24.22$\pm$ 0.06\\
{\it X2$\_1$} & 22.84 $\pm$ 0.04 & 22.46 $\pm$ 0.03 & 23.58 $\pm$ 0.03 & 23.64 $\pm$ 0.02 & 23.61 $\pm$ 0.04 & 23.33$\pm$0.20 & 23.56 $\pm$ 0.03 \\
X5$\_1$ & 24.19 $\pm$ 0.51 & 24.37 $\pm$ 0.11 & 25.99 $\pm$ 0.33 & 24.77 $\pm$ 0.08 & 24.85 $\pm$ 0.23 & 24.80$\pm$ 0.43 & 24.74 $\pm$ 0.18\\
{\it X5$\_2$} & 25.27 $\pm$ 0.17 & 24.48 $\pm$ 0.12 & 24.95 $\pm$ 0.16 & ... & ... & ... & ...\\
{\it X6$\_1$} & ... & ... & ... & ... & ... & ... & 25.33 $\pm$ 0.11 \\
{\it X7$\_1$} & 23.75 $\pm$ 0.11 & 24.12 $\pm$ 0.08 & 25.72 $\pm$ 0.10 & 24.63 $\pm$ 0.07 & 24.78 $\pm$ 0.12 & 24.95$\pm$ 0.65 & 24.97 $\pm$ 0.13\\
{\it X7$\_2$} & ... & ... & ... & ... & 25.84 $\pm$ 0.85 & 24.40$\pm$ 0.41 & 24.12 $\pm$ 0.13\\
\hline
\multicolumn{7}{c}{Flux (10$^{-18}$ (erg cm$^{-2}$ s$^{-1}$ Å$^{-1}$)} \\
\hline
{\it X1$\_1$} & 4.20 $\pm$ 0.05 & 2.88 $\pm$ 0.05 & 0.86 $\pm$ 0.01 & 1.75 $\pm$ 0.03 & 0.91 $\pm$ 0.01 & 0.61 $\pm$0.34 & 0.30 $\pm$ 0.02 \\
{\it X1$\_2$} & ... & ... & ... & ... & ... &...&0.23 $\pm$ 0.06 \\
{\it X2$\_1$} & 3.37 $\pm$ 0.03 & 3.07 $\pm$ 0.04 & 1.47 $\pm$ 0.03 & 2.22 $\pm$ 0.02 & 1.19 $\pm$ 0.04 & 0.83 $\pm$ 0.10 & 0.43 $\pm$ 0.03 \\
{\it X5$\_1$} & 0.88 $\pm$ 0.05 & 0.58 $\pm$ 0.01 & 0.16 $\pm$ 0.06 & 0.78 $\pm$ 0.01 & 0.40 $\pm$ 0.10 & 0.21 $\pm$ 0.07 & 0.14 $\pm$ 0.02 \\
{\it X5$\_2$} & 0.33 $\pm$ 0.02 & 0.52 $\pm$ 0.01 & 0.42 $\pm$ 0.02 & ... & ...& 0.04 $\pm$ 0.01 & 0.04 $\pm$ 0.01 \\
{\it X6$\_1$} & ... & ... & ... & ... & ... & ... & 0.08 $\pm$ 0.02 \\
{\it X7$\_1$} & 1.18 $\pm$ 0.11 & 0.73 $\pm$ 0.01 & 0.20 $\pm$ 0.01 & 0.89 $\pm$ 0.01 & 0.44 $\pm$ 0.01 & 0.19 $\pm$ 0.09 & 0.12 $\pm$ 0.01 \\
{\it X7$\_2$} & ... & ... & ... & ... & 0.16 $\pm$ 0.09 & 0.30 $\pm$ 0.09 & 0.25 $\pm$ 0.02\\
\hline
\end{tabular}
\label{T:fotometri}
\end{table*}

\begin{table*}
\centering
\begin{minipage}[b]{0.9\linewidth}
\caption{Properties of optical counterparts}
\begin{tabular}{ccccccccccccc}
\hline
Optical counterparts & M$_{V_{ACS}}$ & M$_{V_{UVIS}}$ & (B-V)$_{0}$ & (V-I)$_{0}$ & ST & log(F$_{X}$/F$_{V}$) & $\alpha_{ox}$ \\
& (1) & (2) & (3) & (4) & (5) & (6) & (7) \\
\hline
{\it X1$\_1$} & -7.13 $\pm$ 0.05 & -6.90 $\pm$ 0.4 & -0.03 & -0.080 & B5-A0 & 51 $\pm$ 6 & -0.69$\pm$0.01 \\
{\it X2$\_1$} & -7.45 $\pm$ 0.04 & -7.48 $\pm$ 0.3 & 0.03 & 0.054 & B9-A3 & 21 $\pm$ 3 & -0.66$\pm$0.01 \\
{\it {\it X5$\_1$}} & -6.21 $\pm$ 0.23 & -5.10 $\pm$ 0.35 & -0.08 & 0.11 & B6-A1 & 13 $\pm$ 2 & -0.71$\pm$0.08 \\
{\it X5$\_2$} & ... & -6.11 $\pm$ 0.16 & -0.47 & ... & ... & ... & -0.55$\pm$0.03 \\
{\it X7$\_1$} & -6.16 $\pm$ 0.12 & -5.34 $\pm$ 0.11 & -0.15 & -0.19 & B2-B6 & 11 $\pm$ 1 & -0.83$\pm$0.02 \\
{\it X7$\_2$} & -5.22 $\pm$ 0.85 & ... & ... & ... & ... & 31 $\pm$ 4 & ... \\
\hline
\end{tabular}
\\ Notes: Absolute magnitudes were obtained (1) from ACS/WFC (2) from ACS/UVIS data with adopted distance 16.3 Mpc \citep{2000AJ....119..612D}. (3) and (4) Color values obtained from F435W-F550M (B-V) and F550M-F814W (V-I) filters, respectively. (5) Spectral types {\it estimated from the color values}. (6) The F$_{X}$/F$_{V}$ ratios. (7) X-ray-UV correlation cast in terms of the so-called optical spectral index ($\alpha_{ox}$). \\
\label{T:tab6}
\end{minipage}
\end{table*}

\begin{table*}
\centering
\begin{minipage}[b]{0.9\linewidth}
\caption{The {\it diskbb} model parameters of three ULXs from the {\it Chandra} observation}
\begin{tabular}{c c c c c c c r r r l }
\hline
Source & N$_{\mathrm{H}}$ & N$_{\mathrm{{\it diskbb}}}$ & T$_{\mathrm{in}}$ & F$_{\mathrm{X_{unabs}}}$ & L$_{\mathrm{X}}$ & {\it P}\\
(1) & (2) & (3) & (4) & (5) & (6) & (7)\\
\hline
X1 & $0.08_{-0.02}^{+0.02}$ & $2.94_{-1.56}^{+1.57}$ & $1.50_{-0.02}^{+0.02}$ & $3.10_{-0.14}^{+0.14}$ & $9.84_{-0.46}^{+0.46}$  & 0.73 \\
X2 & $0.01_{-0.01}^{+0.02}$ & $3.30_{-1.78}^{+1.78}$ & $1.28_{-0.02}^{+0.02}$ & $1.81_{-0.10}^{+0.10}$ & $5.76_{-0.31}^{+0.31}$  & 0.63\\
X6 & $0.86_{-0.16}^{+0.22}$ & $1.07_{-2.82}^{+2.82}$ & $1.25_{-0.05}^{+0.05}$ & $0.57_{-0.08}^{+0.08}$ & $1.82_{-0.27}^{+0.27}$  & 0.76\\
\hline
\end{tabular}
\\ Note. — Col. (1): Source label. Col. (2): X-ray absorption value, in units of 10$^{22}cm^{-2}$. Col. (3): Normalization parameter of {\it diskbb} model; 10$^{-3}$$\times$([(r$_{in}$ km$^{-1}$)/(D/10 kpc)]$^{2}$ $\times cosi$). Col. (4): Temperature at inner disk radius (keV). Col. (5): Unabsorbed flux in units of 10$^{-13}$ ergs $cm^{-2}$ $s^{-1}$.
Col. (6): Unabsorbed luminosity in units of 10$^{39}$ ergs $s^{-1}$. Col. (7): The null hypothesis probability, which is the probability of the observed data being drawn from the model given the value of $\chi^2$ and the dof. The reduced $\chi^{2}$ is given in parentheses. All errors are at 90\% confidence level. Unabsorbed flux and luminosity values are calculated in the 0.3–10 keV energy band. Adopted distance of 16.3 Mpc \citep{2000AJ....119..612D} is used for luminosity. \\
\label{T:xmodel}
\end{minipage}
\end{table*}

\section*{Acknowledgements}
\noindent
This research was supported by the Scientific and Technological Research Council of Turkey (TÜBİTAK) through project number 119F315. ES and KD acknowledge support provided by the TÜBİTAK through project number 119F334. 

\section*{Data Availability}
The scientific results reported in this article are based on archival observations made by the {\it Chandra}\footnote{https://cda.harvard.edu/chaser/}, XMM-Newton\footnote{http://nxsa.esac.esa.int/nxsa-web/} and Swift-XRT\footnote{https://www.swift.ac.uk/swift\_portal/} X-ray Observatories. This work has also made use of observations made with the NASA/ESA Hubble Space Telescope, and obtained from the data archive at the Space Telescope Science Institute\footnote{https://mast.stsci.edu/portal/Mashup/Clients/Mast/Portal.html}

\bibliographystyle{mnras}
\bibliography{ngc1672} 

\bsp	
\label{lastpage}
\end{document}